\documentclass[aps,tightenlines,twocolumn,prd,nofootinbib,
superscriptaddress,showpacs,preprintnumbers,eqsecnum,floatfix]{revtex4-2}
\def\bea{\begin{eqnarray}}
\def\be{\begin{equation}}
\def\ee{\end{equation}} 
\def\eea{\end{eqnarray}}

\def\sfrac#1#2{{\textstyle \frac{#1}{#2}}}

\usepackage[dvips,usenames]{color}

\usepackage{graphicx}
\graphicspath{{Figures/}} 
\usepackage{amsmath}
\usepackage{amssymb}
\usepackage{bbold}
\usepackage{bm}
\usepackage{slashed}
\usepackage{float}
\usepackage{psfrag}
\usepackage{footnote} 
\usepackage{xcolor}
\usepackage{textcomp}
\usepackage{cancel}
\usepackage{mathtools}
\usepackage[normalem]{ulem}
\begin{document}
%

\title{Quark mass functions in Minkowski space}

\author{Elmar P. Biernat}
\email{elmar.biernat@tecnico.ulisboa.pt}
\affiliation{Departamento de Engenharia e Ci\^encias Nucleares, Instituto Superior T\'ecnico, Universidade de Lisboa, Campus Tecnol\'ogico e Nuclear, 2695-066 Bobadela, Portugal}
\affiliation{Centro de Ci\^encias e Tecnologias Nucleares (C$^2$\!TN), Instituto Superior T\'ecnico, Universidade de Lisboa, Campus Tecnol\'ogico e Nuclear, 2695-066 Bobadela, Portugal}
\affiliation{Laborat\'orio de Instrumenta\c{c}\~ao e F\'isica Experimental de Part\'iculas (LIP), Avenida Professor Gama Pinto, 2, 1649-003 Lisboa, Portugal}

\author{Franz Gross }
 \affiliation{Theory Center Senior Staff (retired), Thomas Jefferson National Accelerator Facility (JLab), Newport News, Virginia 23606, USA}
 \affiliation{Emeritus Professor of Physics, College of William and Mary, Williamsburg, Virginia 23188,
USA}

 \author{M. T. Pe\~na}
\affiliation{Departamento de F\'isica, Instituto Superior T\'ecnico, Universidade de Lisboa, Avenida Rovisco Pais, 1, 
1049 Lisboa, Portugal}
\affiliation{Departamento de Engenharia e Ci\^encias Nucleares, Instituto Superior T\'ecnico, Universidade de Lisboa, Campus Tecnol\'ogico e Nuclear, 2695-066 Bobadela, Portugal}
 \affiliation{Laborat\'orio de Instrumenta\c{c}\~ao e F\'isica Experimental de Part\'iculas (LIP), Avenida Professor Gama Pinto, 2, 1649-003 Lisboa, Portugal}

\author{Alfred Stadler}
\affiliation{Departamento de F\'isica, Universidade de \'Evora, 7000-671 \'Evora, Portugal}
\affiliation{Laborat\'orio de Instrumenta\c{c}\~ao e F\'isica Experimental de Part\'iculas (LIP), Avenida Professor Gama Pinto, 2, 1649-003 Lisboa, Portugal}
\affiliation{Departamento de F\'isica, Instituto Superior T\'ecnico, Universidade de Lisboa, Avenida Rovisco Pais, 1, 
1049 Lisboa, Portugal}

\date{\today}
 \begin{abstract}
  Using the Covariant Spectator Theory (CST), we calculate the dressed quark mass function and wave-function renormalization for the five quark flavors from up/down to bottom in both the spacelike and timelike regions of Minkowski space. The calculation employs a model dressed-gluon propagator fitted to lattice data in the spacelike region. The remaining free parameters of the quark self-energy are determined by fits to lattice results for the quark mass function including a constraint built into the 
  wave-function renormalization. We perform these fits using lattice data from two different groups and find that the resulting mass predictions are within the range of the typical constituent masses used in quark models.  
\end{abstract}


\maketitle


  \section{Introduction}
  Understanding how the observed masses of hadrons emerge from QCD requires a consistent description of both dressed quarks and nonperturbative gluon dynamics. The present work builds on, and substantially extends, the study of dynamical quark-mass generation initiated in Ref.~I~\cite{Biernat:2018khd} within the Covariant Spectator Theory (CST)~\cite{Gross:1969eb,Gross:1982}.
  
 Traditionally, the quark self-energy and quark masses are obtained by solving the Dyson--Schwinger equation (DSE) for the dressed quark propagator in Euclidean space~\cite{Gross:2022hyw}. Recent studies have investigated the analytic continuation of such results to Minkowski space~\cite{Alkofer_2024,Salas-Bernardez:2026fot}.
 Other approaches have solved the DSE directly in Minkowski space within the quenched approximation and rainbow-ladder truncation~\cite{Duarte:2022yur}, or have applied dispersive subtraction methods to DSEs to extract Minkowski-space spectral functions~\cite{Sauli:2020dmx}. The present approach uses the CST to calculate the quark self-energy directly in Minkowski space. This allows us to obtain the dressed quark mass function and wave-function renormalization in both the spacelike and timelike regions. To our knowledge, the obtained results provide the first CST predictions of the dressed quark propagator for all five flavors, from up/down to bottom, in entire Minkowski space.

The CST is related to the Bethe--Salpeter/Dyson--Schwinger (BSDS) framework~\cite{Eichmann:2016bf,PhysRevC.79.012202,Rojas2013,Maas:2011se,Binosi:2009qm,PhysRevD.75.087701,0954-3899-32-8-R02,Maris:2003vk,ALKOFER2001281,Tandy:1997qf,ROBERTS1994477}, from which it can be constructed. At the same time, it may be viewed as a relativistic quark model, since its quark--quark interaction kernel contains a covariant phenomenological generalization of a linear confining potential, supplemented by a one-gluon-exchange (OGE) interaction. The CST effectively goes beyond the ladder approximation commonly employed in BSDS calculations, with well-defined one-body Dirac and non-relativistic Schrödinger limits. A recent overview of the CST approach is given in the volume \textit{50 Years of Quantum Chromodynamics}~\cite{Gross:2022hyw}. For details of the CST in the context of quark mass generation, the reader is referred to Ref.~I.

A calculation of the dressed gluon propagator has not yet been carried out within CST. In previous CST applications, simple phenomenological models of the gluon propagator were used to describe the dressed quark mass function~\cite{Biernat:2014jt,Biernat:2018khd}, as well as the masses of heavy and heavy-light mesons, including highly excited and higher-spin states~\cite{Stadler:2026acl,Biernat:2026owp,Leitao:2017mlx,Leitao:2017it,Leitao2017,Leitao:2014}. In the present work, we take a different step: lattice results for the gluon dressing function~\cite{Aguilar:2010gm}, together with lattice results for the quark mass functions and wave-function renormalizations~\cite{Bowman:2005vx,Oliveira:2018lln}, are used as input for a dynamical calculation of the quark self-energy.

The results of this work have two immediate consequences. First, consistency between the gluon propagator used in the quark self-energy and that used in quark--antiquark bound-state calculations will require a new generation of CST meson studies. Second, because the present work addresses the nonperturbative gluon propagator, it opens a natural connection to current investigations of the gluonic sector of QCD, including possible glueball states~\cite{besiiicollaboration2026lightest0glueballdominant,Nature2026Glueball}.

This paper introduces several important improvements over Ref.~I. Some of these differences are summarized in Appendix~\ref{App:B} and will not be discussed in detail in the main text. One important difference is that the gluon propagator used here depends only on the invariant four-momentum squared. Although this restriction may not be strictly necessary, it makes the simultaneous implementation of Lorentz invariance and chiral symmetry breaking comparatively straightforward within CST.

The implementation of chiral symmetry in CST was developed by Gross and Milana~\cite{Gross:1991te,Gross:1991pk,GMilana:1994} in the early 1990s, and later by Gross and \c{S}avkl\i{}~\cite{Savkli:1999me}. We have investigated this issue further in more recent work~\cite{Biernat:2014jt,pionff:2014,pionff:2015,Biernat:2014xaa,Biernat:2012ig}, deriving constraints on the Lorentz structure of the linear-confining part of the interaction kernel. These constraints allow the linear-confining kernel to decouple from dynamical quark mass generation, leading to a substantial simplification that is exploited in the construction used here.

The paper is organized as follows. Section~\ref{sec:I} presents the general definitions of the dressed quark propagator in CST, introduces the phenomenological quark form factor, and shows the dressed-mass results obtained when the gluon interaction is replaced by a constant interaction. Section~\ref{sec:II} presents the calculation of the dressed OGE contribution. The concluding discussion is given in Sec.~\ref{sec:IV}.

\section{Review of the formalism} \label{sec:I}

\subsection{The dressed propagator}

The  dressed quark propagator $S(p)$ for a bare (current) quark mass $m_0$ and four-momentum $p$, is obtained from the non-linear equation
\bea
S(p)= S_{0}(p)- S_{0} (p) \Sigma(\slashed {p}) S(p)\, ,
\label{eq:DE}
\eea
where  
\bea
S_0(p)=\frac1{Z_0}\frac{m_0+\slashed{p}}{m_0^2-p^2- \mathrm{i}\,\epsilon}\, ,
\eea
is the bare quark propagator with $Z_0$, which depends on $m_0$, canceling the renormalization of the wave function at the bare pole $p^2=m_0^2$,
and 
\bea
\Sigma(\slashed{p})= A(p^2)+\slashed{p}\, B(p^2) \label{eq:selfe}
\eea	
is the quark self-energy, expressed in terms of the renormalized  structure functions $A$ and $B$. The treatment of renormalization in this paper differs substantially from that in Ref.~I, but most of the other discussion is similar.

\begin{figure}[t]
 \centering
 \includegraphics[width=3 in]{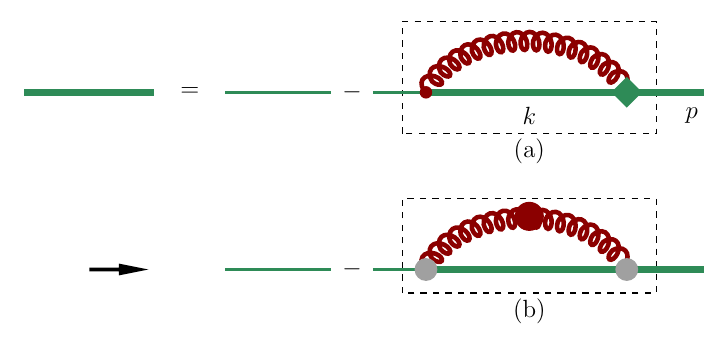}
 \caption{(Color online) A diagrammatic illustration of Eq.~(\ref{eq:DE}).  The diagrams (a) and (b) defined by the dashed boxes are representations of the self-energy $\Sigma(\slashed{p})$.}
 \label{fig:DSE}
\end{figure}
The diagrammatic form of Eq.~(\ref{eq:DE}) is  illustrated in Fig.~\ref{fig:DSE}.  The thin green lines represent $S_0$, the thick green lines $S$.  In box (a), the thick red gluon line is the dressed gluon propagator, the red dot is the bare quark-gluon vertex, and the green diamond is the dressed quark-gluon vertex.  (If the dressed vertex and propagators are exact, they include all of the dressings so that (a) is the exact QCD expression for the quark self energy, see, for instance, Ref.~\cite{ROBERTS1994477}.) 
In this paper, the exact self-energy (a) is approximated by box (b) consisting of a phenomenological dressed gluon propagator, shown as the thick red gluon line with the red blob, which couples to the quarks with a constant $\alpha_{\rm s}$, dressed by a phenomenological form factor at the quark-gluon vertex (shown as gray blobs). The dressed gluon propagator will be taken from lattice calculations.

The solution of Eq.~(\ref{eq:DE}) is then  
\begin{eqnarray}
S(p)&=&\frac1{Z_0(m_0-\slashed{p})+\Sigma(\slashed{p})- \mathrm i\,\epsilon}
=
\frac{Z (p^2) [M(p^2)+ \slashed{p}]}{M^2(p^2)-p^2-\mathrm i\,\epsilon}\,, \nonumber\\  \label{eq:quarkprop}
\end{eqnarray}
where $Z(p^2)$ is the quark wave-function renormalization function and $M(p^2)$ the quark mass function,  related to the invariant self-energy functions $A(p^2)$ and $B(p^2)$ by  
\begin{eqnarray} 
	Z(p^2)&=&\frac{1}{Z_0-B(p^2)}\label{eq:Z} \, , 
    \\
	M(p^2)&=&Z(p^2) \left[ Z_0 m_0+A(p^2) \right] \, , \label{eq:M}
\end{eqnarray}  
respectively.
One of the central assumptions of the CST is that the dressed quark propagator has a real mass pole at $p^2=m^2$, such  that the mass function satisfies 
\begin{eqnarray} 
M(m^2)&=&m \, , \label{eq:const_quark_mass}
\end{eqnarray}
which defines the dressed (constituent) quark mass $m$. 

It should be stressed that the existence of this pole of a single on-shell quark is not in conflict with quark confinement in multiquark systems: in the CST, quark confinement is realized through the special properties of the confining interaction kernel, which is introduced such as to prevent both quark and antiquark in meson states, or all three quarks in baryon states, to be on-shell simultaneously~\cite{Savkli:1999me}.

We require that the self-energy functions $A(p^2)$ and $B(p^2)$ vanish in the asymptotic limit,  
	\bea \lim_{p^2\to \pm \infty}A(p^2)&=& 0\nonumber\\
    \lim_{p^2\to \pm \infty}B(p^2)&=& 0\, \label{eq:asymp}
	\eea 
	 so that
    \begin{subequations}
	\bea \lim_{p^2\to \pm \infty} Z(p^2)&=& \frac1{Z_0}\,,\label{eq:asympZ}\\
	 \lim_{p^2\to \pm\infty} M(p^2)&=& m_0\, . \label{eq:asympM}
	\eea
    \end{subequations}
The function $Z(p^2)$ is renormalized at $p^2=-\mu^2$, with $\mu=3 \text{ GeV}$ and $\mu=1 \text{ GeV}$, corresponding to the renormalization points adopted in Refs.~\cite{Bowman:2005vx} and~\cite{Oliveira:2018lln}, respectively, from which the lattice data used as input in this work will be taken.

Expanding the dressed propagator around $p^2=m^2$ and keeping only leading terms of order $(m^2-p^2)^{-1}$, gives 
\bea
S(p)&\to&\frac{Z(m^2)(m+\slashed{p})}{(m^2-p^2)[1-2mM'(m^2)]}\nonumber\\
&=&\frac{Z_m(m+\slashed{p})}{m^2-p^2-\mathrm i\,\epsilon}
\eea
where
\begin{eqnarray}
&&M'(m^2)=\frac{\mathrm d M(p^2)}{\mathrm d p^2}\bigg|_{p^2=m^2}\,\,\nonumber\\ &&Z_m=\frac{Z(m^2)}{1-2m M'(m^2)}\qquad\;\;\label{eq:Zm}
\end{eqnarray}
where $Z_m$ differs from $Z(m^2)$ because of corrections to the residue coming from the momentum dependence of $M(p^2)$. 
Note that $Z_m$ depends on the dressed quark mass $m$ and is thus flavor-dependent.  
 
 \begin{figure}[t]
 \centering
 \includegraphics[width=3.4 in]{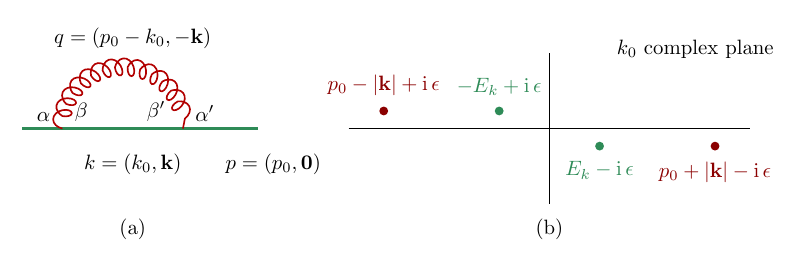}
  \caption{(Color online) (a) Skeleton diagram representing Eq.~(\ref{eq:DEk00}) with momenta and Dirac indices labeled.  (b) Illustration of the $k_0$ complex plane showing location of the quark and gluon poles for a fixed $|{\bf k}|$. } 
 \label{fig:CIT1}
\end{figure}

\subsection{Dynamical calculation of the self-energy}

In the BSDS formalism, using the approximation shown in Fig.~\ref{fig:DSE}(b), the self-energy is calculated from the interaction  of the phenomenological dressed gluon propagator---hereafter referred to simply as the gluon propagator, which is identical to the $q\bar{q}$ interaction kernel---with the dressed quark propagator,
\bea
\Sigma_{\alpha\alpha'} (\slashed p)= -\mathrm i\,h_m^2(p^2)\int \frac{\mathrm d^4k}{(2\pi)^4} \mathcal{V}_{\alpha\beta,\alpha'\beta'} (q^2)  S_{\beta'\beta}(k)\, ,  
\nonumber\\
\label{eq:DEk00}
\eea 
where $h_m(p^2)$ is a quark form factor associated with each quark line that regularizes the CST, 
the quark Dirac indices are shown explicitly, and it is assumed that the kernel ${\cal V}$ depends on $q^2=(p-k)^2$ only (this is more restrictive than the assumptions used in Ref.~I).  The skeleton diagram for $\Sigma$ is illustrated in the left panel of Fig.~\ref{fig:CIT1}.

 In the CST, one is instructed to carry out the $\mathrm d^4k$ integration ignoring the singularities of the kernel at $k_0=p_0\pm (|{\bf k}| - \mathrm i\, \epsilon)$, and keep only the pole from the dressed quark propagator.  If we close the $k_0$ contour in the lower half plane we get the pole at $k_0= E_k - \mathrm i\, \epsilon$; in the upper half plane the pole at  $k_0= -E_k + \mathrm i\, \epsilon$. The contributions from the positive and negative energy poles, which are on the opposite sides of the real $k_0$ axis, pinch when $m\to0$, giving an unwanted singularity. In order to maintain a smooth limit as the quark mass goes to zero, and avoid this singularity, the {\it average\/} over the two contours is taken.  This is necessary for a charge-conjugation symmetric formulation of the CST ~\cite{Savkli:1999me}.
 
 Averaging over these two poles gives the following result \cite{Biernat:2018khd} for the self-energy of a quark with dressed mass $m$:
 \begin{eqnarray}
\Sigma_{\alpha\alpha'} (\slashed p)&=& \sfrac12 Z_m\,h_{m}^2(p^2)\sum_{\sigma=\pm}\int_{\bf k}\mathcal V_{\alpha\beta,\alpha'\beta'} (\hat {q}^2_\sigma)  \Lambda_{\beta'\beta} (\hat k_\sigma ) \, .\nonumber\\
\label{eq:DEk01}
\end{eqnarray} 
 The integrand depends on the internal quark four-momentum  
$\hat k_\sigma=(\sigma E_k,{\bf k})$, where \mbox{$E_k=\sqrt{m^2+{\bf k}^2}$}, with $\sigma=+$ holding for the positive-energy pole term and $\sigma=-$ for the negative-energy pole term.  The equation is covariant, so it may be  evaluated in the  rest frame of the external quark, $p=(p_0,{\bf 0})$, where the integrand simplifies and depends on $E_k$ only.  In this case, the integration reduces to
 \begin{eqnarray}
\int_{\bf k}& \equiv & \int  \frac{\mathrm d^3 {\bf k}}{(2\pi)^3} \frac{m}{E_k}\to \frac{m}{4\pi^2}\int_{m^2}^\infty \frac{k}{E_k} \mathrm dE_k^2 \nonumber\\
&=&\frac{m}{2\pi^2}\int_{m}^\infty k \,\mathrm dE_k  \, . 
\label{eq:kint}
\end{eqnarray}
The interaction kernel, $\mathcal V (\hat q_\sigma^2)$, will  be specified below, and
\begin{eqnarray}
 \Lambda (\hat k_\sigma)=\frac{m+\hat {\slashed{k}}_\sigma}{2m}\,  \label{eq:Lam}
\end{eqnarray}
is the operator projecting onto positive ($\sigma=+1$) or negative ($\sigma=-1$) quark energy states. 
 
The momentum transfer in the rest frame  is
 \bea
\hat q^2_\sigma &=&(\hat k_\sigma-p)^2  =m^2 +p_0^2-2\sigma p_0  E_k  \nonumber\\
&=&m^2+p_0^2-2\sigma \sqrt{y}
\, ,\quad\label{eq:qsigma}
 \eea 
 where $y=p_0^2 E_k^2\to (p\cdot \hat k_\sigma)^2$.

\subsection{The hadronic quark form factor}
  
    The hadronic form factor depends on the flavor (mass) but is otherwise universal (i.e., the same for all interactions involving the same flavored quark).  It is usually removed from the propagator and incorporated into the interaction kernel, with $h_m(p^2)$ attached to the interaction at each end of the line (and multiplication by the same factor for each external line).  In self-energy calculations, where the incoming  and outgoing four-momenta are always the same, an overall factor of $h^2_m(p^2)$ emerges, and the internal loop over ${\hat k}_\sigma$ in (\ref{eq:DEk01})  contributes with the term $h^2_m(m^2)$, which we will normalize to unity.  
    
In this paper, the hadronic quark form factor  (which is purely phenomenological) will be given the form
\bea
h_m^2(p^2)=\frac{m^2\sqrt{\lambda_1^2+1} }{\sqrt{\lambda_1^2 m^4+p^4}}\, , \label{eq:hath}
\eea
where $\lambda_1$ is a dimensionless parameter independent of the quark mass $m$, so that the form factor will depend only on the ratio $p^2/m^2$.  
This form factor suppresses the large $p^2$ behavior, providing convergence for two-body equations (but not, as it turns out, for the self-energies).  The choice of a square root in the definition (\ref{eq:hath}) gives
\bea
\lim_{p^2\to\infty}h_m^2\left(p^2\right)=\sqrt{\lambda_1^2+1}\,\frac{m^2}{p^2} \, ,\label{eq:oneoverp2}
\eea
ensuring the $1/p^2$ behavior needed to fit the lattice data at space-like momenta~\cite{ Bowman_2002,Bowman_2003}.

\subsection{Self-energy predictions from a covariant constant kernel} \label{sec:3}

To begin with, in this subsection we assume a constant kernel. The reason is that the constant interaction plays a major role in removing the singular contributions that arise in the self-energy functions when these are calculated with the OGE interaction. We will see in Subsection~\ref{secIIIE} that, with an appropriate choice of coefficients $C_{\rm s}$ and $C_{\rm v}$ (defined below), only finite, renormalized self-energy functions remain.

In the external quark rest frame, a Lorentz-vector-exchange constant kernel with scalar and vector components is 
\begin{eqnarray}
{\cal V}_{\rm c}&=& {\cal N}_c
\,(2\pi)^3\frac{E_k}{2m}\delta^3({\bf k})  \gamma_\mu(C_{\rm s}+\slashed{\hat{q}}_\sigma C_{\rm v})\otimes\gamma_\nu \qquad
\nonumber\\&&
\times \, \left[\mathrm g^{\mu\nu}-(1-\xi)\frac{\hat q_\sigma^\mu \hat q_\sigma^\nu}{\hat q_\sigma^2}\right] , \quad\label{eq:Ckernel}
\end{eqnarray} 
where $\xi$ is the gauge parameter, with $\xi=0, 1, $ and 3  corresponding to Landau, Feynman-'t Hooft, and Yennie gauges, respectively, 
and the color factor is
\bea
{\cal N}_c=\left[\frac3{16}\sum_a\lambda_a\otimes \lambda_a\right] = 1\, ,  \label{eq:Nc}
\eea

\noindent with $\lambda_a$'s the usual Gell-Mann matrices of $\mathrm {SU}(3)_\text{color}$.
 This sum is unity for both quark self-energies and $q\bar q$ color-singlet interactions.  Note that both the color factor and gauge structure are modeled after the OGE contributions discussed below. The constants $C_{\rm s}$ and $C_{\rm v}$ are the unrenormalized strengths of the interaction; $C_{\rm s}$ differs from the one used in Ref.~I; $C_{\rm v}$ has not been previously discussed.  The factor of 1/2, omitted from Ref.~I, is a renormalization to account for the average over the two channels ($\sigma=\pm1$).  Substituting this operator into Eq.~(\ref{eq:DEk01}) and using the first form of the integral (\ref{eq:kint}) gives 
\bea
&&\hspace{-0.2in}\frac{8\Sigma(\slashed{p})}{{ Z_m} h_m^2(p^2)}=\sum_\sigma \Big[\gamma^\mu \{C_{\rm s}+C_{\rm v}(\sigma m-p_0)\gamma^0\}(1+\sigma\gamma^0)\gamma_\mu\nonumber\\
&&\qquad-(1-\xi)\gamma^0\{C_{\rm s}+C_{\rm v}(\sigma m-p_0)\gamma^0\}(1+\sigma\gamma^0)\gamma^0)\Big]\nonumber\\
&&\qquad=(6+2\xi)(C_{\rm s}+m C_{\rm v}) +(6-2\xi) C_{\rm v}p_0\gamma^0\,.\quad
\eea
Hence the structure functions for a constant interaction are
\bea
A_{\rm c}(p^2)&=&\frac14(3+\xi)( C_{\rm s}+mC_{\rm v})\,Z_m h_m^2(p^2)\nonumber\\
B_{\rm c}(p^2)&=&\frac14(3-\xi)C_{\rm v}\,Z_m h_m^2(p^2)\,  .
\eea

    The physical origin of the constant kernel is interpreted in this paper as a phenomenological description of higher order self-energy contributions  which do {\it not\/} confine.   Any confining parts with a vector structure may be included in the confining part of the kernel, discussed in Refs.~\cite{Leitao:2017mlx,Savkli:1999me}.

\begin{figure}[H]
 \includegraphics[height=2.1in]{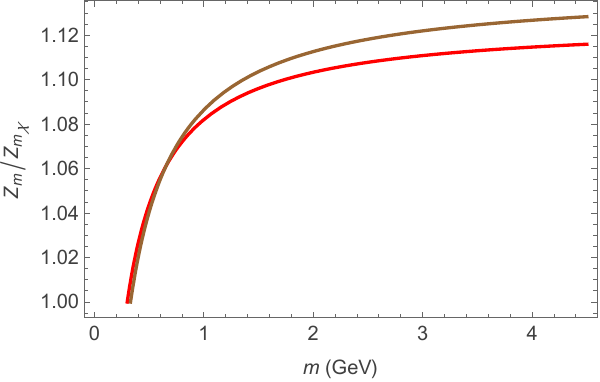} 
  \caption{(Color online) The ratio $Z_m/Z_{m_\chi}$ as a function of $m$ for two parameter sets:
$m_\chi=300~\mathrm{MeV}$ and $\lambda_1=3.85$, corresponding to the
lower curve at large $m$ shown in red, and $m_\chi=330~\mathrm{MeV}$ and
$\lambda_1=3.62$, corresponding to the upper curve at large $m$ shown in
brown. }
 \label{fig:Zm}
\end{figure}

In the following, we show that, interestingly, even in the absence of the OGE interaction, the constant interaction alone, with a suitable choice of the quark form factor, provides a very good description of the lattice data for the quark mass function. 
If we choose $C_{\rm v}=0$ and $Z_0=1$,  and define $C\equiv\sfrac14(3+\xi)C_{\rm s}$ to be independent of $\xi$, the mass equation (\ref{eq:M}) becomes
\bea
M(p^2)=m_0+C\, Z_mh_m^2(p^2)\, .\label{eq:Mc}
\eea
At the mass shell point $p^2=m^2$, the normalization of $h$ gives 
\bea
m=m_0+CZ_m\, . \label{eq:massm}
\eea
In the chiral limit, $m_0=0$, the gap equation then fixes $C$,
\bea 
C =\frac{m_\chi}{Z_{m_\chi}} \, ,
\label{eq:barC}
\eea  
where $m_\chi$ is the dressed quark mass in the chiral limit. The mass function
in the chiral limit is  
\bea
M_\chi(p^2)
&=& A_\chi(p^2)= m_\chi h_{m_\chi}^2(p^2)=m_\chi\sqrt{\frac{\lambda_1^2+1}{\lambda_1^2+{\displaystyle \frac{p^4}{m^4_\chi}}}}\, \nonumber\\ 
\, \label{eq:MC}
\eea
 and Eq.~(\ref{eq:Zm}) can be solved analytically,
\bea
Z_{m_\chi}=\frac{1+\lambda_1^2}{3+\lambda_1^2}\,,
\eea
and thus
\bea
C=m_\chi\frac{3+\lambda_1^2}{1+\lambda_1^2}\, .
\eea  
For finite bare mass, Eq.~(\ref{eq:Zm}) has two solutions,
\bea
Z_m^\pm=\frac{2\sqrt{\lambda_1^2+1}}{\sqrt{\lambda_1^2+1}\pm\sqrt{1+\lambda_1^2 +\frac{m_\chi}{m}\frac{16}{\lambda_1^2+1}+8\frac{m_\chi}{m} }}\,.\nonumber\\
\eea
The physical solution is the one with the plus sign in the denominator. This branch has the correct limiting behavior: it approaches 1 in the limit $\lambda_1\to \infty$ and reduces to the chiral-limit value $Z_{m_\chi}$ when $m\to m_\chi$. 
\noindent The mass function for a dressed quark of mass $m$ is then predicted to be
\bea 
M_m(p^2)=m_0+m_\chi \frac{Z_{m}}{Z_{m_\chi}}\sqrt{\frac{\lambda_1^2+1}{\lambda_1^2+{\displaystyle \frac{p^4}{m^4}}}}\, .\label{eq:Mc1}
\eea
This is consistent with the simple prediction in Eq.~(\ref{eq:massm}).  Since $Z_m$ itself depends on the quark mass $m$, Eq.~(\ref{eq:massm}) is, in general, nonlinear. However, in the present model, the dependence $m$ of $Z_m$ is rather mild over the entire range considered, from the chiral-limit value to the bottom-quark masses, as illustrated in Fig.~\ref{fig:Zm}.
Thus, to keep the model simple, we approximate $Z_m$ by expanding it about $m_\chi/m=1$ and keeping only the leading-order contribution. 
The mass function then reduces to
\bea 
M_m(p^2)\simeq m_0+m_\chi \sqrt{\frac{\lambda_1^2+1}{\lambda_1^2+{\displaystyle \frac{p^4}{m^4}}}}\, \qquad\label{eq:Mc2}
\eea
which is similar to the ``simple'' Ansatz in Refs.~\cite{Bowman_2002,Bowman_2003} for $\alpha=1$.

From each of the two lattice-data references, Refs.~\cite{Bowman:2005vx}
and~\cite{Oliveira:2018lln}, we selected four data sets corresponding
to different lattice quark masses and performed combined fits using
Eq.~\eqref{eq:Mc2}. In these fits, we fixed the chiral-limit mass
$m_\chi$ and adjusted one global parameter, $\lambda_1$, together with
four bare quark masses, $m_{0}^i$, where the index $i=1,\ldots,4$ labels the lattice data set.

It is emphasized that the reason the $m_{0}^i$'s must be fit to the lattice data is that the values of the bare quark mass specified in the lattice calculations (that is $m_{\rm lattice}^i=16, 32, 47,$ and $63$ MeV for Ref.~\cite{Bowman:2005vx} and  $m_{\rm lattice}^i=6.2, 8, 17,$ and $18.4$ MeV for Ref.~\cite{Oliveira:2018lln}) are renormalized by lattice effects at small distances, so that the final $m_{0}^i$'s that determine the asymptotic values of the mass functions are not the same as the lattice input numbers.

Each lattice data set for the mass function
associated with a lattice bare-quark mass $
m_{\rm lattice}^i$ contains 261 data points for Ref.~\cite{Bowman:2005vx} and 78 data points for Ref.~\cite{Oliveira:2018lln}. In the combined fits to the mass-function lattice data we determine the global parameter
$\lambda_1$ and the four bare quark masses $m_0^i$, one for each $m_{\rm lattice}^i$. Therefore, the combined fit to the data of Ref.~\cite{Bowman:2005vx} uses 1044 data points in total, while the combined fit to the data of Ref.~\cite{Oliveira:2018lln} uses 312 data points in total. 

\begin{figure*}
 \centering
 \leftline{\includegraphics[width=3.2in]{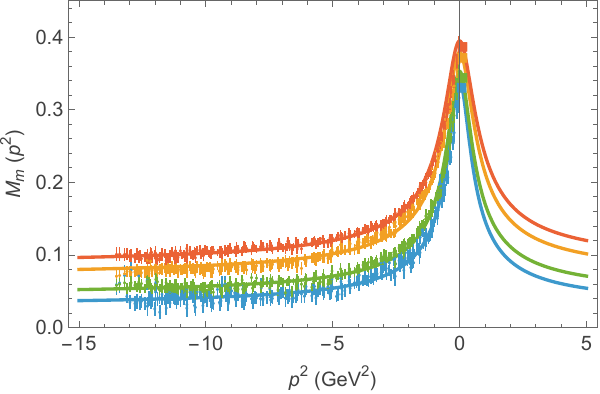}}
 \vspace*{-2.1in}
\rightline{\includegraphics[width=3.2in]{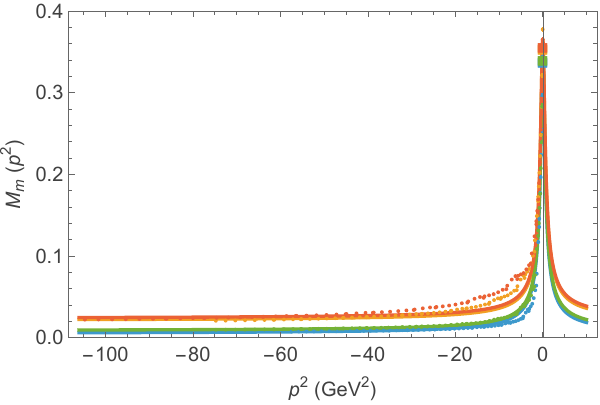}}  
 \leftline{\includegraphics[width=3.2in]{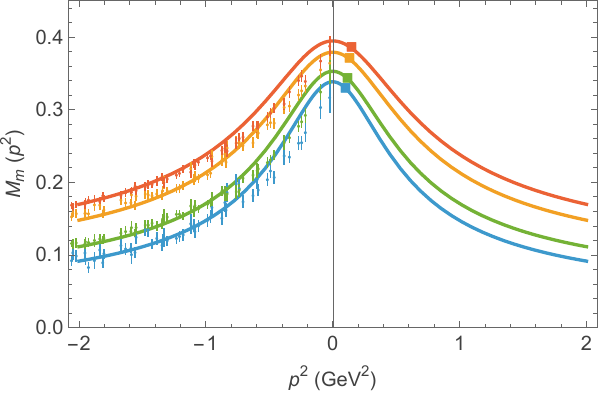}}
 \vspace*{-2.1in}
\rightline{\includegraphics[width=3.2in]{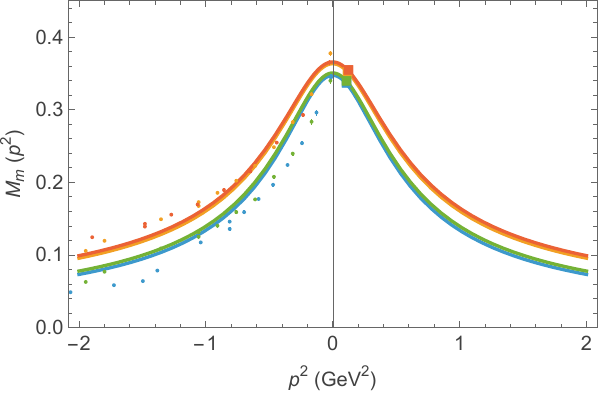}}  
\hspace{0.2in}
  \caption{(Color online) 
 Mass functions obtained with the constant kernel of Eq.~(\ref{eq:Ckernel}). The left panels show the combined fit to the lattice data of Ref.~\cite{Bowman:2005vx}, while the right panels show the combined fit to the lattice data of Ref.~\cite{Oliveira:2018lln}. Each colored curve corresponds to the lattice data shown in the same color, as specified in Table~\ref{tab:masses}. From bottom to top, the curves correspond to increasing quark mass. In the fits, the chiral quark mass was set to $m_\chi= 300$ MeV for the data of Ref.~\cite{Bowman:2005vx} and to $m_\chi= 330$ MeV for the data of Ref.~\cite{Oliveira:2018lln}. The large dots indicate the points $\{m^2,m\}$ for all four cases.}
 \label{fig:Mc}
\end{figure*}

The predicted dressed-quark masses obtained with this simple model are listed in Table~\ref{tab:masses}, together with those from our best OGE model, which is determined in the following sections. The corresponding mass functions are compared in Fig.~\ref{fig:Mc} with lattice data from the two independent studies, Refs.~\cite{Bowman:2005vx,Oliveira:2018lln}. The agreement with the data of Ref.~\cite{Bowman:2005vx} is very good, indicating that the simple model performs well for this data set. By contrast, applying the same procedure to the data of Ref.~\cite{Oliveira:2018lln} leads to a significantly poorer description, pointing to a possible discrepancy between the two lattice data sets.

\section{Self-energy from the OGE kernel}  \label{sec:II}

In this section, lattice data is used to fix the parameters of models describing a dressed OGE kernel, and to predict the mass function for quarks of different bare masses.   

The major issue that arises here is that a straightforward application of the CST to these self-energy calculations gives singularities at $p^2 = 0$. 
These singularities can be removed using a new normalization technique defined and described in Sec.~\ref{sec:Dresults}.
\subsection{Formulae for $A$ and $B$}
In the Feynman-’t Hooft gauge, the effective OGE from Ref.~I is
\begin{eqnarray}
\mathcal V_{\rm g}(q^2)=\frac{16\pi\alpha_0}{3} {\cal N}_c D (q^2) \;\gamma_\mu\otimes\gamma^\mu
 \, , \qquad
  \label{eq:OGEgauge}
\end{eqnarray}
where  ${\alpha}_0$ is an adjustable parameter and the color factor ${\cal N}_c = 1$ was defined in Eq.~(\ref{eq:Nc}) above.
The $1/(-q^2)$ which would appear for a massless OGE is replaced by the function $D (q^2)$ which models the $q^2$ dependence of a dressed OGE.
The renormalization method used in this paper eliminates the need for the form factor $g(y)$ used in Ref.~I.

To find the equations for the self-energy, substitute (\ref{eq:Lam}) and (\ref{eq:OGEgauge}) into (\ref{eq:DEk01}) and  carry out the Dirac algebra in the quark rest frame:  
\bea
\gamma^\mu\Big[\frac{m+\hat{\slashed{k}}_\sigma}{2m}\Big]\gamma_\mu &=& 2-\frac{\hat{\slashed{k}}_\sigma}{m}\to 2-\frac{\sigma E_k}{m}\gamma^0\, ,
\label{eq:reduction}
\eea
in agreement with Eq.~(3.19) of Ref.~I.  This gives  
  \begin{eqnarray}
\Sigma (\slashed p)= \frac{8\pi\,  Z_m\alpha_0}{3} \,h_m^2(p^2)\sum_{\sigma=\pm}\int_{\bf k}D  (\hat{q}^2_{\sigma\pm})\Big[2-\frac{\sigma E_k}{p^0}\frac{\slashed{p}}{m}\Big]\, .\nonumber\\
\label{eq:DEk03}
\end{eqnarray}
Correcting some errors, including a factor of $1/(2m)$ omitted from Eqs.~(3.13) and (3.14) of Ref.~I (but included correctly in subsequent equations), and using the integral reduction (\ref{eq:kint}), the results for $A$ and $B$ now are 
\begin{eqnarray}
\frac{{A} (p^2)}{h^2_{m}(p^2)}&\equiv& m\overline{A} (p^2)
= m\frac{ 8 Z_m\alpha_0}{3\pi}\int_{m}^\infty k \,\mathrm d E_\mathrm k
\sum_\sigma D (q_\sigma^2)
\, ,
\nonumber\\
\frac{{B}(p^2)}{h^2_{m}(p^2)}&\equiv&\overline{B}(p^2)
=-\frac{4 Z_m\alpha_0}{3 \pi} \int_{m}^\infty k \, \mathrm d E_k  \sum_\sigma
\frac{E_k}{p_0}\sigma D(q_\sigma^2)\, .\nonumber\\
\label{eq:Feynman}
\end{eqnarray}
Note that both  $\overline{A}$ and $\overline{B}$ are dimensionless, but $A$ has dimensions of mass. 
For the time being, it is assumed that $p^2\geq0$, so $p_0$ and $\sqrt{y}$, from Eq.~(\ref{eq:qsigma}), are real. 

In Sec.~\ref{sec:Mass} below, the renormalized versions of $\overline{A}$ and $\overline{B}$ will be fit to the lattice mass functions directly, implying that the form factor $h^2_{m}$ in Eq.~(\ref{eq:Feynman}) is unity.

\begin{figure*}
 \leftline{\includegraphics[height=2.1in]{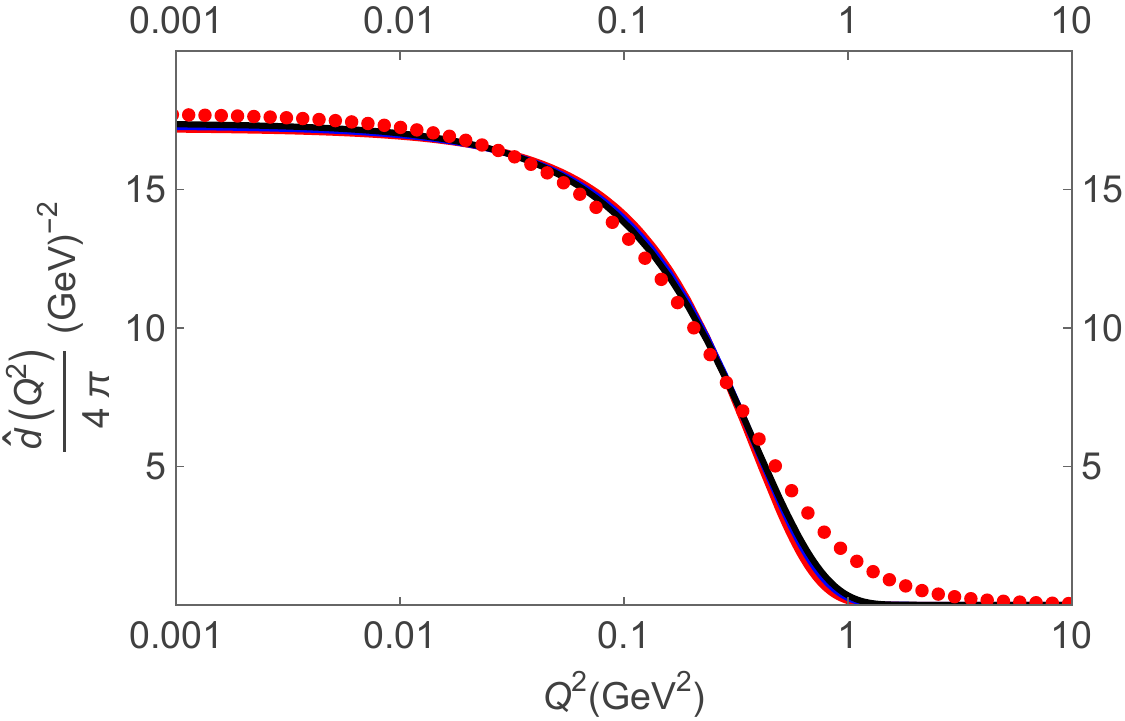}}
 \vspace*{-2.1in}
\rightline{\includegraphics[height=2.1in]{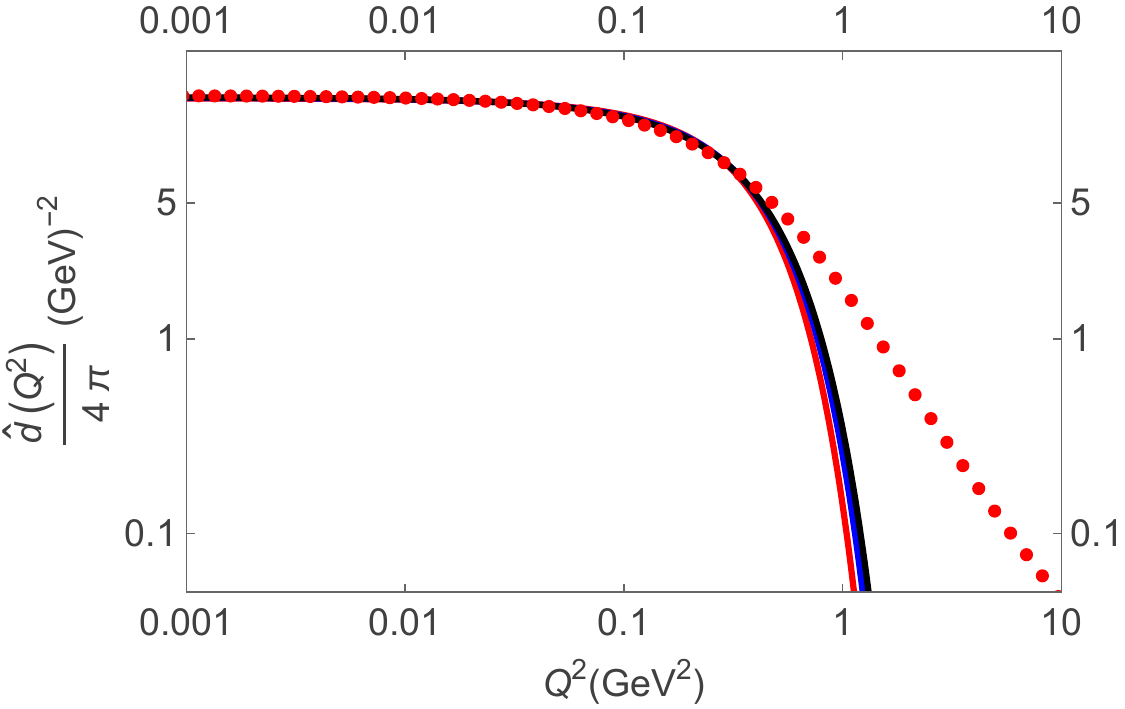}}  
  \caption{(Color online) Lattice data for $\widehat{d}(Q^2)$ from Ref.~\cite{Aguilar:2010gm} (red dots) compared to lines showing the fits defined in Table \ref{tab:fitpar}.  The differences between the three models are hardly visible in either panel. }
 \label{fig:Gg}
\end{figure*}

\subsection{Models for the dressed gluon propagator}
\begin{table}[b]
\begin{minipage}{2.7 in}
 \caption{Model parameters for $D(Q^2)$.    
Parameter values shown in boldface were kept fixed during the fit.
}

\begin{ruledtabular}
 \begin{tabular}{lcccr}
Model &  $\gamma$ & $M_1$ (GeV) & $\alpha_0$ & Line color\\[0.05in]
\hline
$ 1$  & {\bf 2}  &  {0.79}    &{17.2}  &   Red     \\[0.05in]
$ 2$  & {\bf  2.25}  &  {0.85}    &{17.3}   & Blue      \\[0.05in]
$ 3$  & {\bf  2.5}  &  {0.90}    &{17.4} & Black         \\[0.05in]
\end{tabular}
\end{ruledtabular}
\label{tab:fitpar} 
\end{minipage}
\end{table}

We have investigated a large number of ways to model the nonperturbative gluon propagator.  We first chose models of the form 
\bea
\alpha_0\,D_1(q^2) &=& \frac{2\alpha_0\,M_1^6}{\big[M_1^8+(M_1^2 -q^2)^4\big]}
\, , \qquad\label{eq:D00}
\eea
 where $\alpha_0$ is the effective QCD fine structure constant appropriate for the dressed gluon exchange  and $M_1$ is a mass parameter.  Note that $D_1$ has no singularities along the entire $q^2$ axis.  These models were stimulated by discussion of the lattice work  of Aguilar, Binosi and Papavassiliou~\cite{Aguilar:2010gm}.  Using $Q^2=-q^2$, they compute the product 
 \bea
 \frac{\widehat{d}(Q^2)}{4 \pi} &=&\frac{\alpha_{\rm s}(Q^2){\cal G}(Q^2)}{Q^2} = \frac{\alpha_{\rm s}(Q^2)}{M_{\rm g}^2(Q^2)+Q^2}\nonumber\\
 &=&\alpha_0 D(Q^2)  \, ,
 \eea 
 where the notation ${\cal G}(Q^2)$ and $M_{\rm g}(Q^2)$ is from Pennington and Wilson \cite{Pennington:2011xs},  Eq.~(13), and denotes the dressing of the gluon propagator which removes the singularity of the bare gluon propagator at $Q^2=0$.  
\begin{figure}
\includegraphics[height=2.1in]{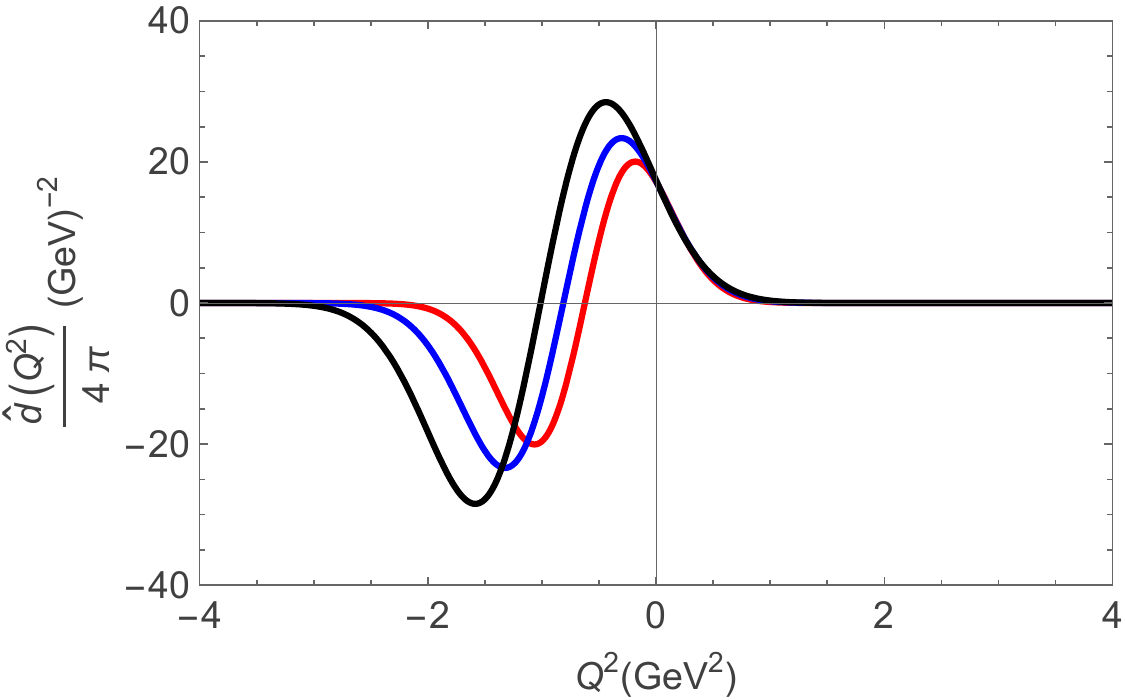}
 \caption{(Color online) Comparison of the three models over the full range of $Q^2$, with the color code specified in Table~\ref{tab:fitpar}. From
right to left, the curve with the first, second, and third maximum corresponds to Model 1,
2, and 3, respectively. The lattice data of Ref.~\cite{Aguilar:2010gm}
(red dots) are available only in the space-like region and are very small
on this scale. The models show significant differences in the time-like
region.}
\label{fig:D00}
\end{figure}

 However, it turns out that substituting models of the type (\ref{eq:D00})  into  Eq.~(\ref{eq:Feynman}) leads to results with singularities at particular values of $m$. 
To avoid these problems, we turned  finally to models of the form 
 \bea
 \alpha_0 D(q^2)=\alpha_0\Big(1-\frac{q^2}{\beta M_1^2}\Big)\exp\left[\frac{\gamma q^2}{M_1^2}-\frac{q^4}{M_1^4}\right]\, ,
 \label{expmodel}
 \eea
with parameters $M_1$, $\beta$, $\gamma$ and $\alpha_0$ constrained by a fit to the lattice data and the asymptotic conditions (\ref{eq:asymp}). In particular, it is shown in the Appendix that the condition
  $$\qquad\qquad\qquad\qquad\qquad\qquad \beta=\frac{\gamma}{2} \nonumber \qquad\qquad\qquad\qquad\text{(\ref{eq:gamma})}$$ 
  
 \noindent ensures that the asymptotic conditions (\ref{eq:asymp}) are satisfied. The remaining free parameters are $\alpha_0$, $\gamma$ and $M_1$.

 Depending on the parameter values, the mass equation (\ref{eq:const_quark_mass}) may have no solution for certain values of $m_0$, or may even have multiple solutions. Three choices of $\gamma$, listed in Table \ref{tab:fitpar}, were studied.   
 As shown in Fig.~\ref{fig:Gg}, all models fit the low-$Q^2$ subset of the 200 lattice data points in the spacelike $q^2$ region quite well, with differences that are hardly visible. However, Fig.~\ref{fig:D00} reveals that their timelike behavior differs substantially. While they agree in the perturbative regime (large momentum transfer), they diverge in the infrared timelike region. Here, a wiggle in the gluon propagator signals complex singularities: self-interactions become massive, and the propagator exhibits a dynamical mass scale with non-perturbative dressings forcing the poles off the real axis.

\begin{figure*}
\leftline{\includegraphics[height=2.1in]{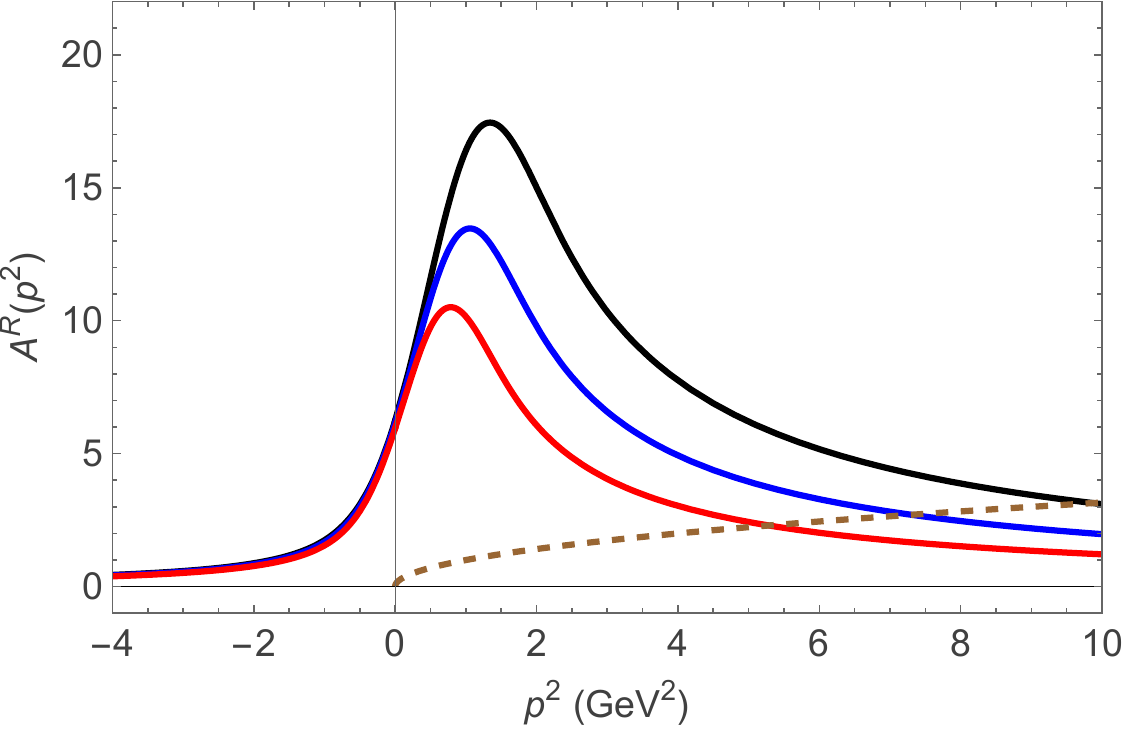}}
\vspace{-2.1in}
\rightline{\includegraphics[height=2.1in]{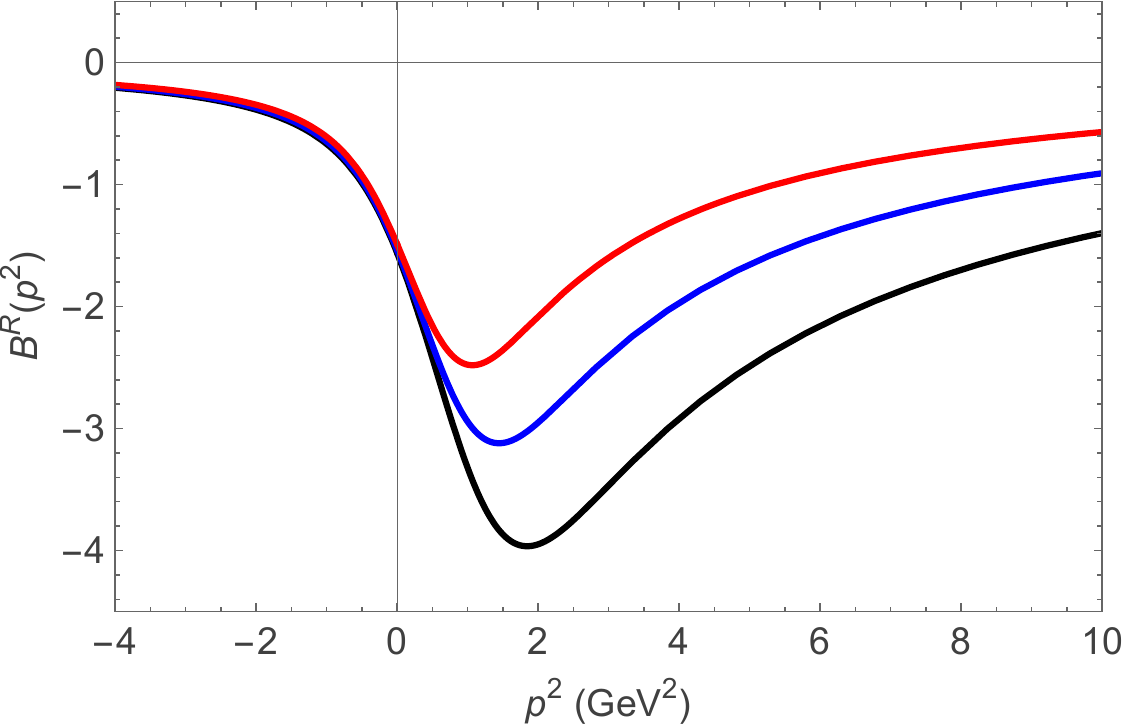}}
   \caption{(Color online) The regularized functions $A^{\rm R}$ (left panel) and $B^{\rm R}$ (right panel)
for the three models, with the color code specified in
Table~\ref{tab:fitpar}. In both panels, the curves from bottom to top correspond to
Models~1, 2, and 3, respectively.  In the left panel, the brown dashed line denotes
$\sqrt{p^2}$. In all cases shown, $m=0.3~\mathrm{GeV}$
and $Z_m=1$.
  }
 \label{fig:ABR}
\end{figure*}

\begin{figure*}
\leftline{\includegraphics[height=2in]{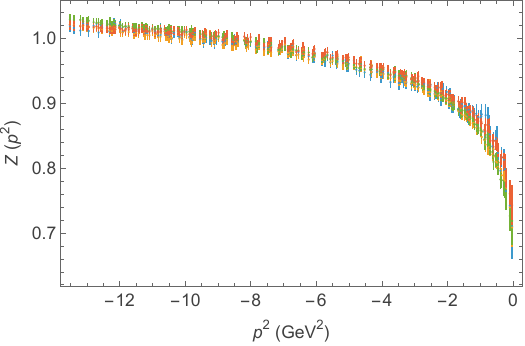}}
\vspace{-2in}
\rightline{\includegraphics[height=2in]{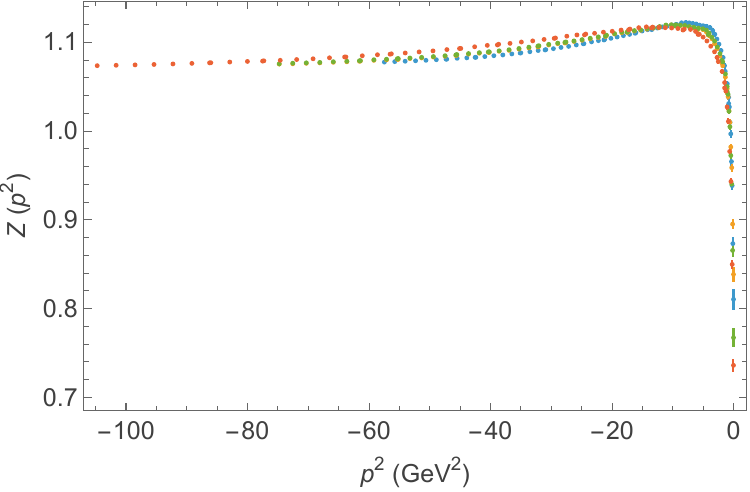}}
 \caption{(Color online) Lattice wave function renormalization data for the 4 lattice cases of \cite{Bowman:2005vx} renormalized at $p^2=-9$ GeV$^2$ (left panel), and that of \cite{Oliveira:2018lln} renormalized at $p^2=-1$ GeV${}^2$ (right panel).}
\label{fig:Zchi}
\end{figure*}

\subsection{Regularization} \label{sec:Dresults}

The integrals (\ref{eq:Feynman}) {are singular} at $p^2=0$, {but are otherwise finite and real everywhere}.  {The regularization procedure will require the functions be  {\it finite \/} at $p^2=0$.}  

Part of the singularity is a pole that can be explicitly displayed by changing variables from  $E_k$ to $y=p^2E_k^2$ (first introduced following Eq.~(\ref{eq:qsigma}) above).  When applying this transformation, it is essential to distinguish the time-like region, $p^2>0$ (where $y>0$) from the spacelike region $p^2<0$ (where $y<0$).  Using $\mathrm dy \to 2p^2 E_k \mathrm d E_k$, the integrals in the time-like region are 
\bea
\overline{A}(p^2)
 &=& \frac{ 8 Z_m\alpha_0}{3\pi\,p^2}\int_{m^2p^2}^\infty \mathrm d y\sqrt{\frac{y-m^2 p^2}{y}} {\cal A}(y,p^2)
  \nonumber\\
\overline{B}(p^2)
 &=& -\frac{4 Z_m\alpha_0}{3 \pi\,p^2}\int_{m^2p^2}^\infty \mathrm d y\sqrt{\frac{y-m^2 p^2}{y}}{\cal B}(y,p^2)\, , 
 \label{eq:AandBR}
\eea
where the kernels are
\bea
{\cal A}(y,p^2)&=&D(\hat{q}_+^2)+D(\hat{q}_-^2) \, ,\nonumber\\
{\cal B}(y,p^2)&=&\frac{\sqrt{y}}{p^2}\big[D(\hat{q}_+^2)-D(\hat{q}_-^2)\big]\, .
\label{eq:A&B}
\eea
While a pole {at $p^2=0$} is nicely displayed by this transformation, there is an additional singularity arising from the square root in the weight function of the integrals.

It is convenient to introduce ``smooth'' contributions where these singularities are removed and only the pole is retained:
\bea
A^{\rm S}(p^2)&\equiv&\frac{8Z_m \alpha_0}{3\pi\,p^2}\int_{0}^\infty \mathrm dy\, {\cal A}(y,p^2) \, ,\nonumber\\
B^{\rm S}(p^2)&\equiv&-\frac{ 4Z_m\alpha_0}{3\pi\,p^2}\int_{0}^\infty \mathrm dy\, {\cal B}(y,p^2) \, .\label{eq:smooth}
\eea 
Letting $F$ represent either $A$ or $B$, the differences between the smooth contributions and the original results, written collectively as
\bea
\delta F(p^2) \equiv \overline{F}(p^2)-F^{\rm S}(p^2)\, , \label{eq:DZi}
\eea
are constructed from integrands localized  around $p^2=0$, and are small.  As they will be removed by renormalization, they will not be discussed further.

It turns out that the  smooth terms (\ref{eq:smooth}) can be integrated analytically. Assuming $p_0>0$, the integrals are analytic functions of $p_0^2 \to p^2$, and can therefore be extended to $p^2<0$.  They can be written
\bea
F^{\rm S}(p^2)=\frac{Z_m\alpha_0}{3 \pi p^2}f^{\rm S}(p^2) \, ,\label{eq:smooth2}
\eea  
where the $f$ functions ($ a$ and $ b$) are given in Eqs.~(\ref{eq:ai}) and (\ref{eq:bi}).

The pole at $p^2=0$ can be removed from the smooth functions by a polynomial subtraction, which will then give a finite, continuous result.  The weight function $a^{\rm S}(p^2)$ is finite as $p^2\to 0$ so the needed subtraction is simply $a^{\rm S}(0)$.
However, $b^{\rm S}(p^2)$ has a pole as $p^2\to 0$, so it must be expanded in a series and the terms of order $(p^2)^{-1}$ and $(p^2)^{0}$ removed.  These details are shown in Appendix \ref{App:A}. The regularized results for $A^{\rm R}$ and $B^{\rm R}$ are shown in Fig.~\ref{fig:ABR} for the three models of Table~\ref{tab:fitpar}.

\subsection{Quark Mass Functions} \label{sec:Mass}

\begin{figure*}
\leftline{\includegraphics[height=2in]{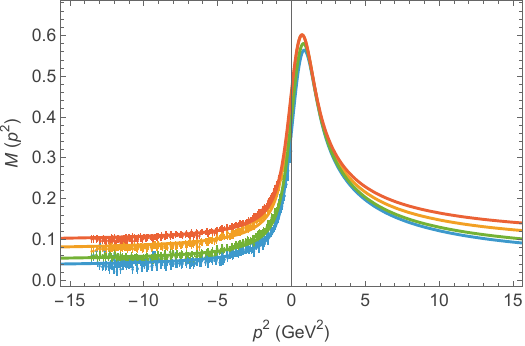}}
\vspace{-2in}
\rightline{\includegraphics[height=2in]{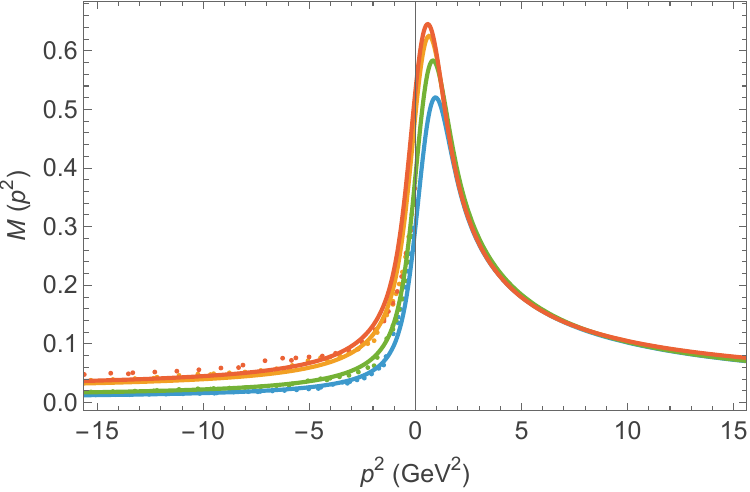}}
\leftline{\includegraphics[height=2in]{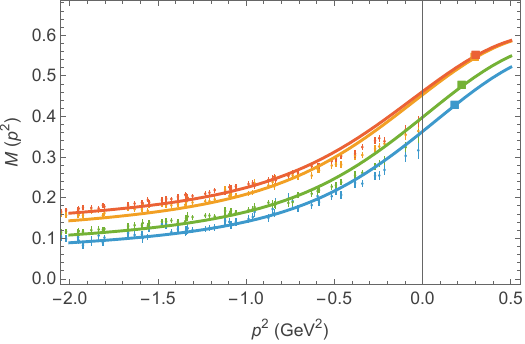}}
\vspace{-2in}
\rightline{\includegraphics[height=2in]{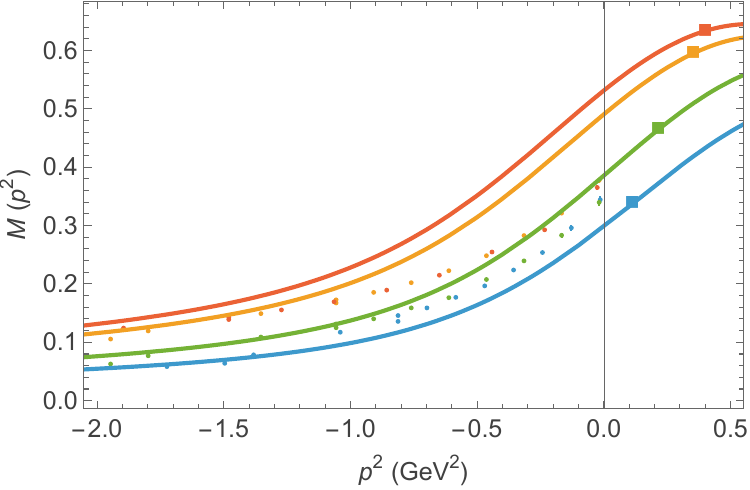}}
\leftline{\includegraphics[height=2in]{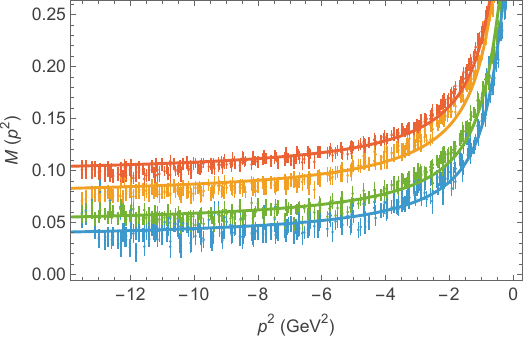}}
\vspace{-2in}
\rightline{\includegraphics[height=2in]{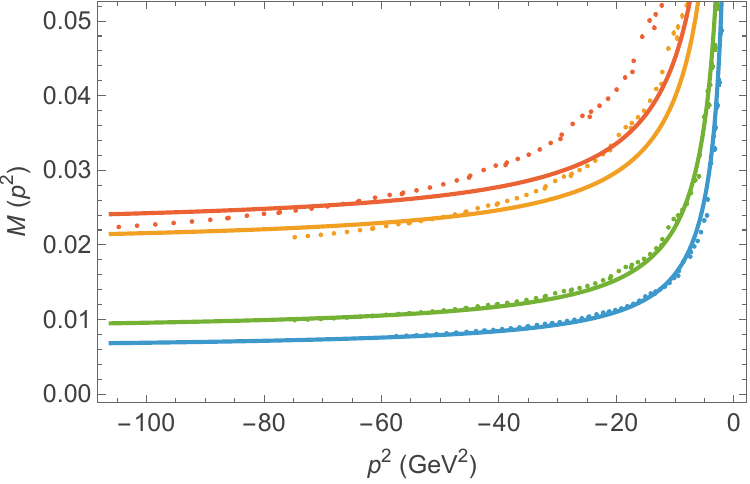}}
 \caption{(Color online) Fits of the Model 2 mass functions to the lattice data of Ref.~\cite{Bowman:2005vx} are shown in the left panels, while the corresponding fits to the data of Ref.~\cite{Oliveira:2018lln} are shown in the right panels. 
In each panel, the curves, from bottom to top, correspond to increasing quark mass, with the color code specified in
Table~\ref{tab:fitpar}. The colored squares, visible in the center panels, indicate the positions of the mass points $(m^2,m)$. The model describes the data of Ref.~\cite{Bowman:2005vx} and two lightest lattice-mass cases of Ref.~\cite{Oliveira:2018lln} very well. For the two heavier lattice-mass cases of Ref.~\cite{Oliveira:2018lln}, the agreement is considerably poorer.} 
\label{fig:allM}
\end{figure*}


	\begin{figure*}
  \begin{minipage}{0.48\textwidth}
    \includegraphics[height=1.7in]{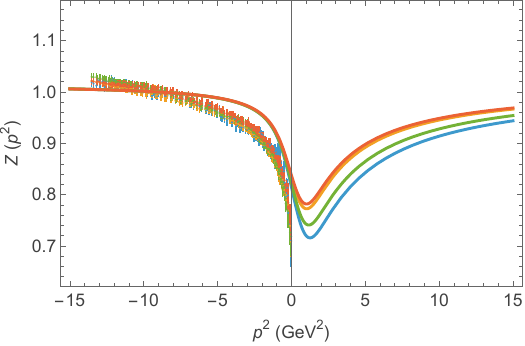}
  \end{minipage}
  \hspace{0.025\textwidth}
  \begin{minipage}{0.48\textwidth}
    \includegraphics[height=1.7in]{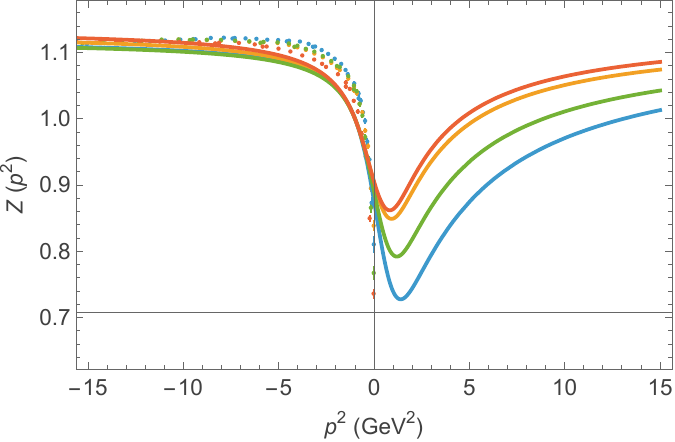}
  \end{minipage}
 \caption{(Color online) The model 2 function $Z(p^2)$ compared to data for the 4 lattice cases of Ref.~\cite{Bowman:2005vx} (left panel) and of Ref.~\cite{Oliveira:2018lln} (right panel). In each panel, the curves, from bottom to top, correspond to increasing quark mass, with the color code specified in
Table~\ref{tab:masses}. Note that the theoretical predictions are similar, tracking the lattice results.}
\label{fig:Zall}
\end{figure*}

\begin{figure}[h]
	\includegraphics[height=1.6in]{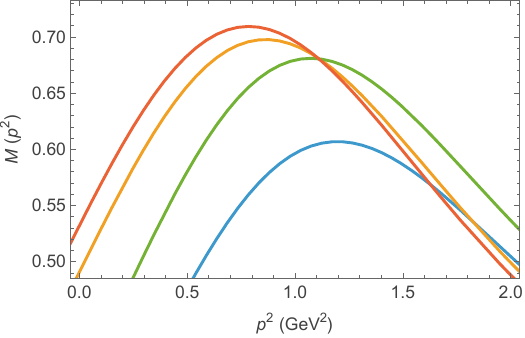}
 \caption{(Color online) Intersections of the Model~3 mass-function curves in the timelike region for the fit to the lattice data of Ref.~\cite{Oliveira:2018lln}.}
\label{fig:M3_M_O}
\end{figure}

\begin{figure}
 	\includegraphics[height=2.1in]{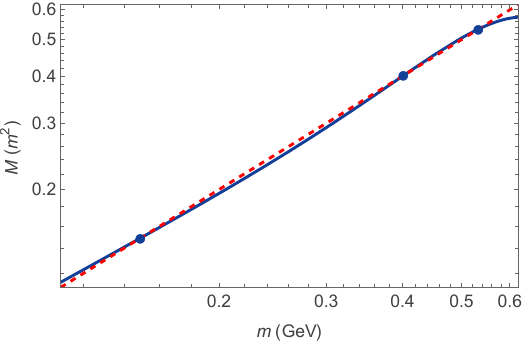}
 	 \caption{(Color online)  The function $M(m^2)$ (dark blue line) obtained from Model~1 fitted to the lattice data of Ref.~\cite{Oliveira:2018lln} for up-quark bare mass. The intersections with the line $M=m$ (red dashed line) are marked by the blue dots showing that there is more than one solution of Eq.~(\ref{eq:masseq}).}
 	\label{fig:M1_Mvsm_O}
 \end{figure}

\begin{figure*}
	\begin{minipage}{0.48\textwidth}
    \includegraphics[height=1.7in]{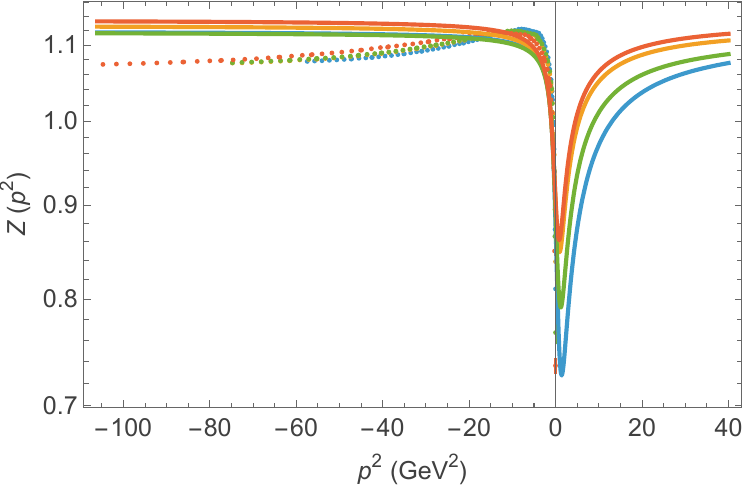}
	\end{minipage}
  \hspace{0.025\textwidth}
  \begin{minipage}{0.48\textwidth}
  \includegraphics[height=1.7in]{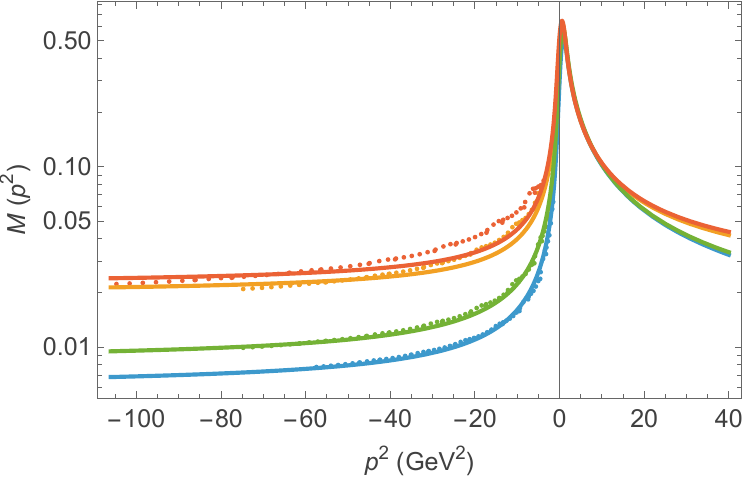}
  \end{minipage}
	 \caption{(Color online) The model 2 function $Z(p^2)$ (left panel) and mass function (right panel), plotted over a large range of $p^2$, are compared to data for the 4 lattice cases of Ref.~\cite{Oliveira:2018lln}. In each panel, the curves, from bottom to top, correspond to increasing quark mass, with the color code specified in Table~\ref{tab:masses}.}
	\label{fig:ZMall_O}
\end{figure*}

\begin{figure*}
\begin{minipage}{0.48\textwidth}
    \includegraphics[height=1.7in]{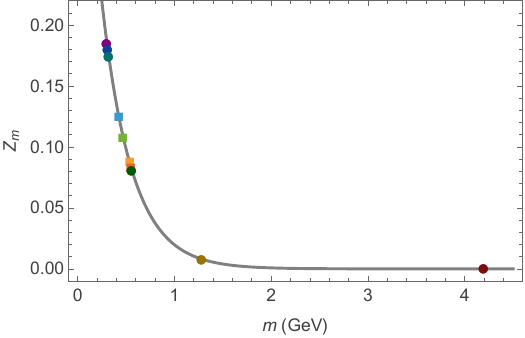}
	\end{minipage}
  \hspace{0.02\textwidth}
  \begin{minipage}{0.48\textwidth}
  \includegraphics[height=1.7in]{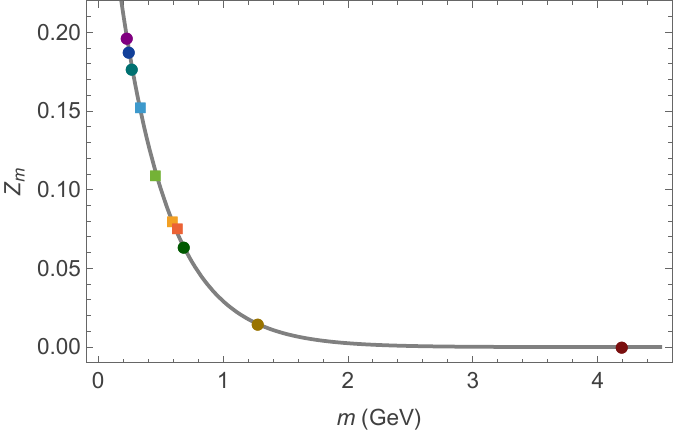}
\end{minipage}
 \caption{(Color online) The fitted function $Z_m$ given in Eq.~(\ref{eq:xivsm}). The squares are the lattice fits to the data of Ref.~\cite{Bowman:2005vx} (left panel) and Ref.~\cite{Oliveira:2018lln} (right panel); the dots are the values obtained for the predicted masses in each case. In each panel, the squares and dots, from top to bottom, correspond to increasing quark mass, with the color code specified in
Table~\ref{tab:masses}.}
\label{fig:xivsm}
\end{figure*}

\begin{figure*}
 	\begin{minipage}{0.48\textwidth}
    \includegraphics[height=1.7in]{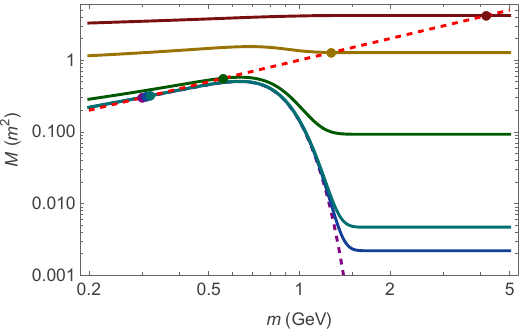}
	\end{minipage}
  \hspace{0.02\textwidth}
  \begin{minipage}{0.48\textwidth}
  \includegraphics[height=1.7in]{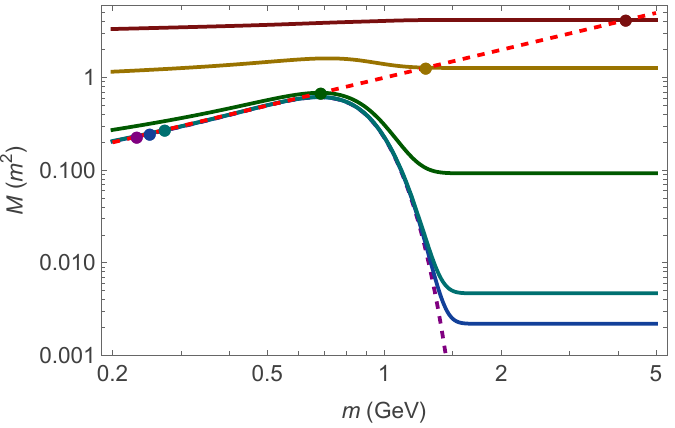}
\end{minipage}
 	 \caption{(Color online)  The function $M(m^2)$ versus $m$ for different flavors, obtained from the fit of Model 2 to lattice data from Refs.~\cite{Bowman:2005vx} (left panel) and~\cite{Oliveira:2018lln} (right panel). In each panel, the curves, from bottom to top, correspond to increasing quark mass, with the color code specified in
Table~\ref{tab:masses}. The intersections with the line $M=m$ (red dashed line) show the locations of the solutions of Eq.~(\ref{eq:masseq}).}
 	\label{fig:Mvsm}
 \end{figure*}


\begin{table}[t]
\begin{minipage}{3.3 in}
 \caption{
 Parameters of the mass-function models given in Eqs.~\eqref{eq:Mc2} and~\eqref{eq:massfnct2}, obtained using the constant interaction and the OGE interaction of Model~2 in Table~\ref{tab:fitpar}, respectively. Bare quark masses ($m_{\rm lattice}$ and $m_0$) and dressed quark masses ($m$) are given in MeV. The bare masses for each flavor are in the \(\overline{\rm MS}\) scheme, as reported by the Particle Data Group~\cite{PDG}. For the constant interaction, we set \(Z_0=1\). For the OGE interaction fitted to the data of Ref.~\cite{Bowman:2005vx}, we imposed \(Z(-\mu^2)=1\) at \(\mu=3~\mathrm{GeV}\), while for the OGE interaction fitted to the data of Ref.~\cite{Oliveira:2018lln}, we imposed the same condition at \(\mu=1~\mathrm{GeV}\). Parameter values shown in boldface were used as inputs for the predictions.}

 \centering 
 \begin{tabular}{l|ccc|ccc|l}
 \hline\hline
 &  \multicolumn{3}{|c|}{Constant} & \multicolumn{3}{|c|}{OGE}& \\[0.05in]

  & $m_0$ &  $m$ & $\lambda_1 $ & $m_0$ &$m$  &  ${Z_m}$ &  Line color\\
 \hline  \hline 
  $m_{\rm lattice}$ &  \multicolumn{6}{|c|}{{\it Fits to data of Ref.~\cite{Bowman:2005vx}}}  & \\[0.03in]
 \hline 
16   &  {28}  &  328     &  {3.85} &{32} &  {427}& {0.124}& Blue  \\[0.05in]
32  &  {43} &  343    &  {3.85}  &  {46 }& {475}& {0.107}& Green  \\[0.05in]
47  & {69}   &  369    &  {3.85}    & {72}& {544}& {0.087}& Orange  \\[0.05in]
63  & {85}   &   385    &  {3.85}   & {94} & {549}& {0.082}& Red  \\[0.05in]
 \hline 
 Flavor  &  \multicolumn{6}{|c|}{{\it Predictions}} &  \\[0.03in]
 \hline 
chiral     & \textbf{{ 0}}          & \textbf{{300}}&  \textbf{{3.85}}  & \textbf{{0}} &{302}& { 0.185}& Dashed, Purple   \\[0.05in]
up      & \textbf{{2.20}} & 302  &  \textbf{{3.85}} & \textbf{{2.20}} &   {311}& {0.180}& Dark Blue\\[0.05in]
down   &\textbf{{4.69}} & 305 &  \textbf{{3.85}}& \textbf{{4.69}} &  {322}&  {0.174}& Dark Cyan\\ [0.05in]
strange & \textbf{{92.7}}& 393&  \textbf{{3.85}} &  \textbf{{92.7}}& {562}&   {0.081}& Dark Green\\ [0.05in]
charmed & \textbf{{1275}} &1575 &  \textbf{{3.85}}& \textbf{{1275}}&{1279}&  {0.008}& Dark Yellow\\ [0.05in]
bottom  & \textbf{{4196}} & 4496 &  \textbf{{3.85}}& \textbf{{4196}} & {4196} &{$10^{-7}$}& Dark Red\\[0.05in] 
		\hline\hline
  $m_{\rm lattice}$&  \multicolumn{6}{|c|}{{\it Fits to data of Ref.~\cite{Oliveira:2018lln}} } & \\[0.03in]
  \hline
		 6.2 &  {5} &335 & {3.62}  & {5.8} &
			{338}& {0.152}
		 & Blue  \\[0.05in]
		8 &  {8.3}  & 338 & {3.62}  & {8.1}  &  
			{465}& {0.108} & Green   \\[0.05in]
		17 &  {21}  & 351 & {3.62}  &   {19.5}  &  
			{596}& {0.079} & Orange  \\[0.05in]
		18.4 & {23.5}  &353 & {3.62}   &  {21.9} &  
			{634} &{0.074}  & Red  \\[0.05in]
    \hline 
 Flavor  &  \multicolumn{6}{|c|}{{\it Predictions}} &  \\[0.03in]
 \hline
		chiral     & \textbf{{0}}         & \textbf{{330}} & \textbf{{3.62}} &\textbf{{0}} & {232}     & {0.196}  &  Dashed, Purple \\[0.05in]
		up      & \textbf{{2.20}} & 332 & \textbf{{3.62}}& \textbf{{2.20}}  &     {250}   & {0.188} & Dark Blue\\[0.05in]
		down   &\textbf{{4.69}} & 335 & \textbf{{3.62}}& \textbf{{4.69}} &  {274}    &  {0.177} & Dark Cyan\\ [0.05in]
		strange & \textbf{{92.7}}& 423 & \textbf{{3.62}}&\textbf{{92.7}} &  {687}  &  { 0.064} & Dark Green\\ [0.05in]
		charmed & \textbf{{1275}} &1605 & \textbf{{3.62}} & \textbf{{1275}}&{1282} &  { 0.015} & Dark Yellow\\ [0.05in]
		bottom  & \textbf{{4196}} & 4526& \textbf{{3.62}} &\textbf{{4196}} &  {4196} &{$10^{-5}$}& Dark Red\\[0.05in] 
          \hline\hline
	\end{tabular}
   
\label{tab:masses} 
\end{minipage}
\end{table}

 From the definition of the mass function given in Eqs.~(\ref{eq:Z}) and (\ref{eq:M}) we obtain

 \bea
M (p^2,m_0)= \frac{Z_0 m_0+m\, A^{\rm R} (p^2)}{Z_0- B^{\rm R} (p^2)}\, . \label{eq:massfnct}
\eea

 Our model for the quark mass function and the wave-function renormalization depends on three global parameters, common to all quark flavors, and four local flavor-dependent fitting parameters assigned separately to each lattice data set for a given lattice bare quark mass.

The first global parameter is $\gamma$, which enters the nonperturbative gluon propagator. Of the three values considered in this work and listed in Table~\ref{tab:fitpar}, only one, corresponding to Model~2 with $\gamma=2.25$, will ultimately be selected on the basis of the requirements discussed below. The other two global parameters in Table~\ref{tab:fitpar} that enter the gluon propagator, $\alpha_{0}$ and $M_{1}$, are fixed by the fit to the lattice gluon-propagator data of Ref.~\cite{Aguilar:2010gm}. They are therefore not free parameters in the fits to the quark mass function.

The first local parameter is $Z_0$, which may equivalently be expressed in terms of $Z(-\mu^2)$, where $\mu$ denotes the renormalization point. This parameter is fixed by imposing the renormalization condition $Z(-\mu^2)=1$. This will be incorporated into the mass and Z functions from the beginning, so will not be explicitly adjusted during the fitting.

The other three local parameters are the bare mass $m_0^i$, the constituent mass $m^i$, and the mass-renormalization factor $Z_m^i$, where the index $i$ labels the lattice data set. The mass function depends on $Z_m$ through $A^{\rm R}$ and $B^{\rm R}$. According to Eq.~(\ref{eq:Zm}), this factor $Z_m$ depends on the wave-function renormalization $Z$ and on the derivative of the mass function, both evaluated at the mass-shell point $p^2=m^2$. Thus, in principle, Eq.~(\ref{eq:Zm}) should be solved self-consistently for $Z_m$ for each flavor. This procedure we leave for future work. In the present study, we adopt a simpler approach and treat $Z_m$ as an additional adjustable local parameter, fitting $Z_m^i$ to each lattice data set.
 
We calibrate our model by fitting separately to the lattice data of
Refs.~\cite{Bowman:2005vx} and~\cite{Oliveira:2018lln} at each value of
the lattice bare quark mass, $m_{\rm lattice}^i$. Once calibrated, the model can be used to calculate mass functions for arbitrary values of $m_0$ without additional assumptions.

The calibration begins with the observation that the wave function renormalization $Z$ for each of the four lattice cases is renormalized at $p^2=-\mu^2$ so that $Z(-\mu^2)=1$, or
\bea
Z_0=1+B ^{\rm R}(-\mu^2)\, .\label{eq:Z0}
\eea
The mass function can therefore be written
\bea
M (p^2,m_0)= \frac{[1+ B^{\rm R} (-\mu^2)] m_0+m\, A^{\rm R} (p^2)}{1+B ^{\rm R}(-\mu^2)- B^{\rm R} (p^2)}\, , \label{eq:massfnct2}
\eea
with
\bea
Z (p^2)= \frac{1}{1+B ^{\rm R}(-\mu^2)- B^{\rm R} (p^2)}\, . \label{eq:Zfnct}
\eea

\noindent The lattice data of Ref.~\cite{Bowman:2005vx} are renormalized at $\mu = 3 \text{ GeV}$, while the lattice data of Ref.~\cite{Oliveira:2018lln} are renormalized at $\mu = 1 \text{ GeV}$.


Using these results, the parameters $m_0$,
$m$, and $Z_m$ are determined, for each model listed in Table~\ref{tab:fitpar}, by separately fitting the lattice mass functions for each $m_{\rm lattice}^i$, subject to the on-mass-shell constraint
\bea
M (m^2,m_0)=m\,.
\label{eq:masseq}
\eea
The four lattice data sets of Ref.~\cite{Bowman:2005vx} correspond to the lattice quark masses
$m_{\rm lattice}^i=16~{\rm MeV}$, $32~{\rm MeV}$, $47~{\rm MeV}$, and $ 63~{\rm MeV},
$
whereas those of Ref.~\cite{Oliveira:2018lln} correspond to
$m_{\rm lattice}^i=6.2~{\rm MeV}$, $8~{\rm MeV}$,  $17~{\rm MeV}$, and $18.4~{\rm MeV}.
$
The parameters were obtained by a 
$
\chi^2
$
minimization carried out with Wolfram Mathematica.

The lattice data for the quark wave-function renormalization are nearly independent of the quark mass, as shown in Fig.~\ref{fig:Zchi}. The four quark-mass cases from both references, displayed with the colors specified in Table~\ref{tab:masses}, overlap almost completely and are therefore difficult to distinguish. This indicates that the choice of
$Z(-\mu^2)$
should be model independent. In the present calibration,
$Z(-\mu^2)=1$
is fixed by Eq.~(\ref{eq:Z0}).

The fits to the mass-function data of Ref.~\cite{Bowman:2005vx} use the 261 data points of each set, whereas the corresponding fits to the data of Ref.~\cite{Oliveira:2018lln} use the 78 data points of each set. For the associated data for the $Z$-function, namely the $4\times 261 = 1044$ points of Ref.~\cite{Bowman:2005vx} and the
$ 4\times 78 = 312$ points of Ref.~\cite{Oliveira:2018lln}, we impose $Z(-\mu^2)=1$
at
$\mu = 3~\mathrm{GeV}$
and
$\mu = 1~\mathrm{GeV}$, respectively, from the beginning. Apart from this normalization condition, the results for
$Z$ are predictions of the mass-function fits.

The results of the fits for Model~2 are reported in Table~\ref{tab:masses}. 
The corresponding mass functions are shown in Figs.~\ref{fig:allM}, the wave-function renormalizations in Fig.~\ref{fig:Zall}.  
These results do not require any quark form factor (that is, $h_m^2(p^2)=1$).

As shown in Fig.~\ref{fig:allM}, Model~2 provides a very good description of the lattice data of Ref.~\cite{Bowman:2005vx}, as well as of the two lower lattice-mass cases of Ref.~\cite{Oliveira:2018lln}. By contrast, the description of the two larger masses reported in Ref.~\cite{Oliveira:2018lln} is poorer, despite the substantially smaller uncertainties of these data compared with those of Ref.~\cite{Bowman:2005vx}. In particular, the corresponding mass functions exhibit more curvature than the data. This raises the question of whether the data sets from the two references are mutually compatible.
We postpone further discussion to the Conclusions.

We also used Models~1 and~3 to fit the lattice data, and both provide a very good description in the spacelike region. However, when Model~3 is fitted to the lattice data of Ref.~\cite{Oliveira:2018lln}, the four resulting mass-function curves cross each other in the timelike region, as shown in Fig.~\ref{fig:M3_M_O}. Consequently, the mass functions corresponding to lower quark masses exceed those corresponding to higher quark masses. We regard this inversion of the expected mass ordering as unphysical. Since this behavior is absent in Models~1 and~2, we discard Model~3, as well as any model with $\gamma\geq 2.5$.

Furthermore, within the CST framework, a physically acceptable model should yield one and only one solution of the mass equation (\ref{eq:masseq}) for each quark flavor. As we will see, Model~2, together with the parametrization of  $Z_m$ given in Eq.~(\ref{eq:xivsm}) below, satisfies this requirement. By contrast, when Model~1 is fitted to the lattice data of Ref.~\cite{Oliveira:2018lln}, more than one solution of the mass equation is obtained for the up-quark mass, as shown in Fig.~\ref{fig:M1_Mvsm_O}. For this reason, Model~1 is rejected.  The fits of Model 2, over a large range of $p^2$, are shown in  Fig.~\ref{fig:ZMall_O}.

 We find that the fitted values of the parameter $Z_m$ listed in Table~\ref{tab:masses} exhibit a strong dependence on the quark mass. We parametrize this flavor dependence using a simple functional form. For Model~2, the fitted expression
\begin{eqnarray}
Z_m= \rho\,  \mathrm{e}^{-\tau\, m}  \label{eq:xivsm}
\end{eqnarray}
provides a good description of the values in Table~\ref{tab:masses}, as shown in Fig.~\ref{fig:xivsm}. The fitted parameters are  $\rho=0.49$, $\tau=3.21$ GeV${}^{-1}$ for the lattice data of Ref.~\cite{Bowman:2005vx}, and  $\rho=0.35$, $\tau=2.48$ GeV${}^{-1}$ for the lattice data of Ref.~\cite{Oliveira:2018lln}.  
With this parametrization, the dressed masses of all quark flavors can be obtained by solving the mass equation~(\ref{eq:masseq}), with all parameters fixed as described above. To visualize the solutions, it is useful to plot the function $M(m^2)$ as a function of $m$, as shown in Fig.~\ref{fig:Mvsm} for Model~2, and to identify its intersections with the line $M=m$. We emphasize that the dependence of $M$ on $m^2$ is not the same as its dependence on $p^2$, as is evident from Eq.~(\ref{eq:massfnct}).

\begin{figure*}
\includegraphics[height=1.9in]{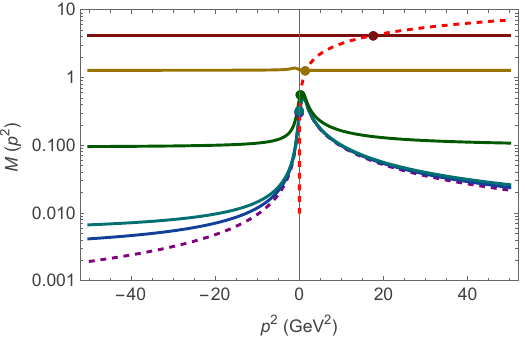}
	\vspace{-2.2in}
	\includegraphics[height=1.9in]{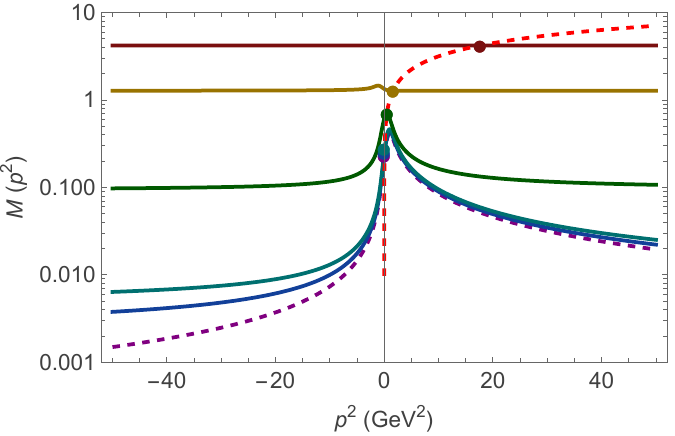}
    \vspace{2.2in}\\
		\includegraphics[height=1.9in]{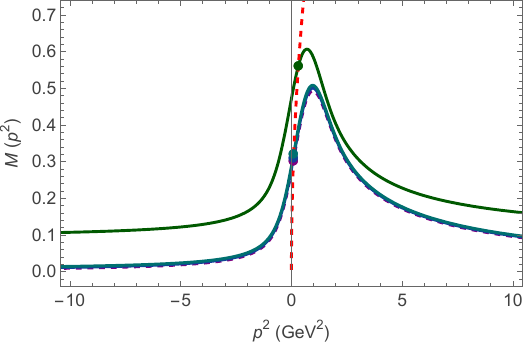}
	\vspace{-2.2in}
	\includegraphics[height=1.9in]{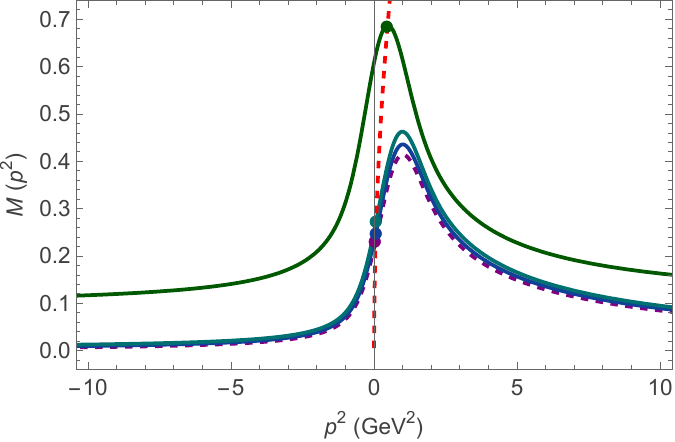}
	\vspace{2.2in}
 \caption{(Color online) Predictions for the Model 2 mass functions from fitting the lattice data of Ref.~\cite{Bowman:2005vx} (left panels) and Ref.~\cite{Oliveira:2018lln} (right panels), in the chiral limit and for all 5 flavors (top panels) and the light and strange quarks (bottom panels). In each panel, the curves, from bottom to top, correspond to increasing quark mass, with the color code specified in
Table~\ref{tab:masses}.
The purple, blue and cyan dots for chiral-limit, up, and down quarks, respectively, are barely distinguishable. The dashed red line in all panels is the mass condition $m = \sqrt{p^2}$. }
\label{fig:massother}
\end{figure*}

\begin{figure*}
	\includegraphics[height=1.9in]{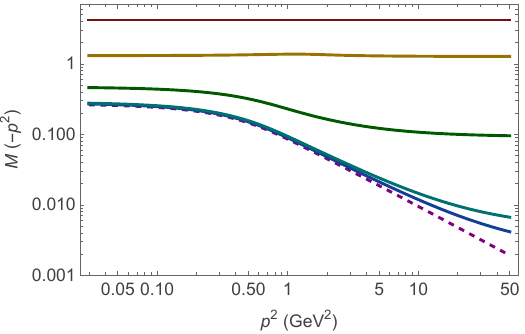}
	\vspace{-2.1in}
    \includegraphics[height=1.9in]{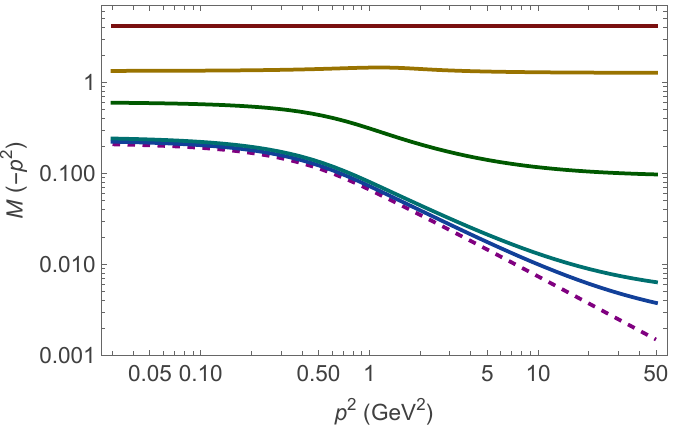}
	\vspace{2.1in}\\
		\includegraphics[height=1.9in]{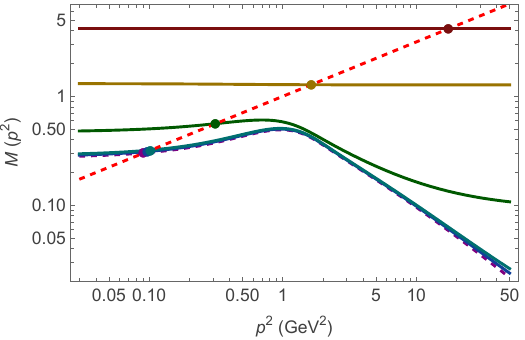}
	\vspace{-2.1in}
	\includegraphics[height=1.9in]{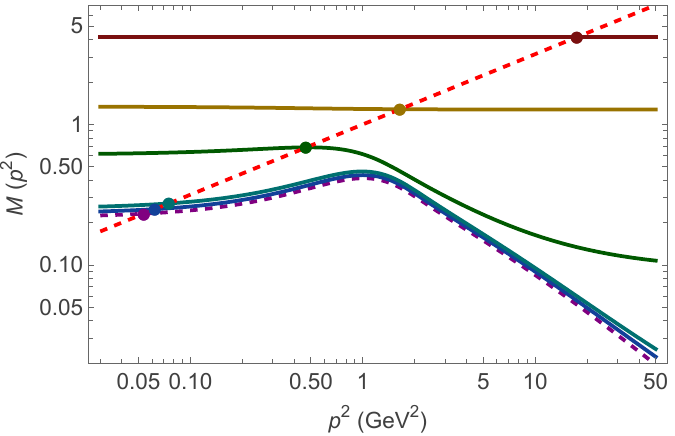}
	\vspace{2.1in}
	 \caption{(Color online) Predictions for the Model 2 mass functions from fitting the lattice data of Ref.~\cite{Bowman:2005vx} (left panels) and Ref.~\cite{Oliveira:2018lln} (right panels), in the chiral limit and for all 5 flavors in the spacelike (top panels) and timelike regions (bottom panels) of Minkowski space. In each panel, the curves, from bottom to top, correspond to increasing quark mass, with the color code specified in
Table~\ref{tab:masses}. The dashed red line in the bottom panels is the mass condition $m = \sqrt{p^2}$. }
	\label{fig:massother2}
\end{figure*}

\begin{figure*}
\includegraphics[height=2.2in]{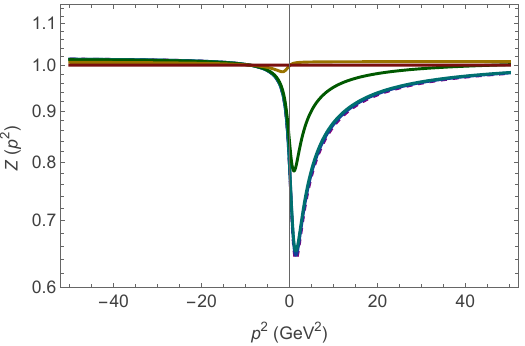}
	\vspace{-2.2in}
    \includegraphics[height=2.2in]{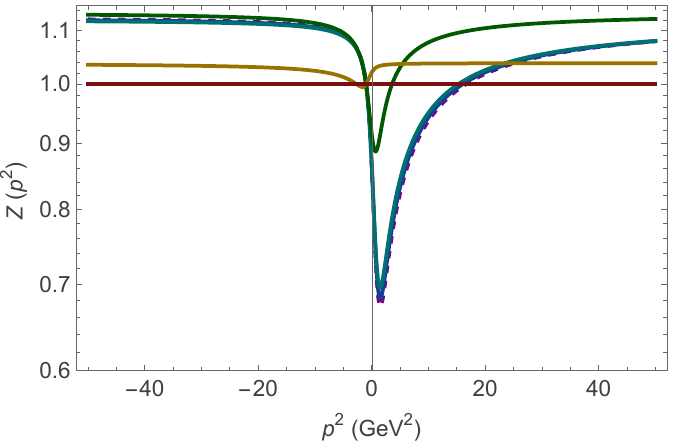}
	\vspace{2.2in}\\
		\includegraphics[height=2.2in]{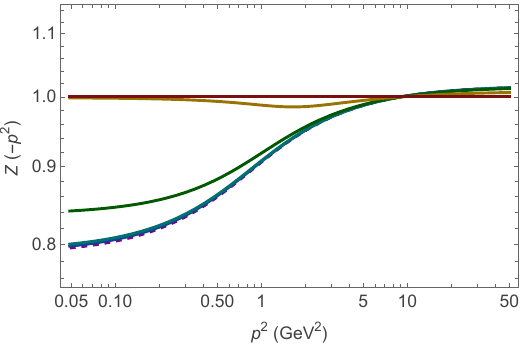}
	\vspace{-2.2in}
	\includegraphics[height=2.2in]{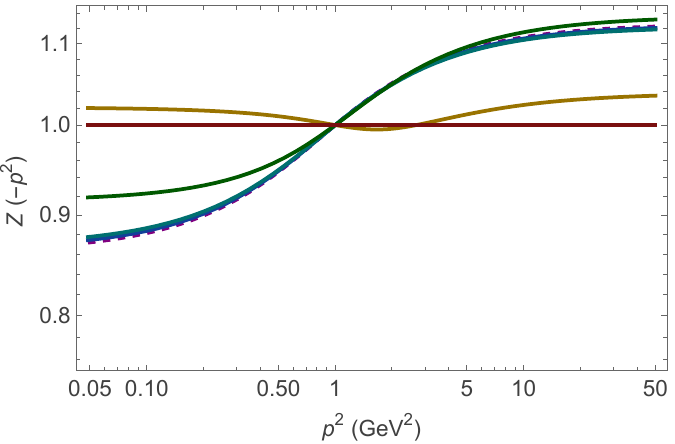}
	\vspace{2.2in}
    \\
		\includegraphics[height=2.2in]{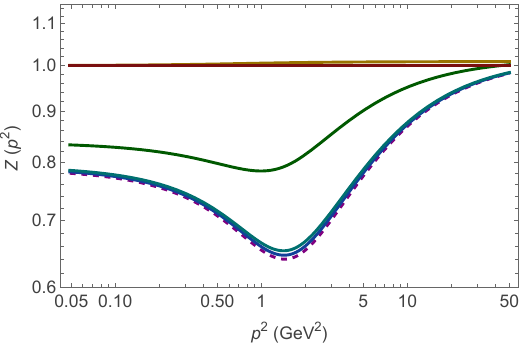}
	\vspace{-2.2in}
	\includegraphics[height=2.2in]{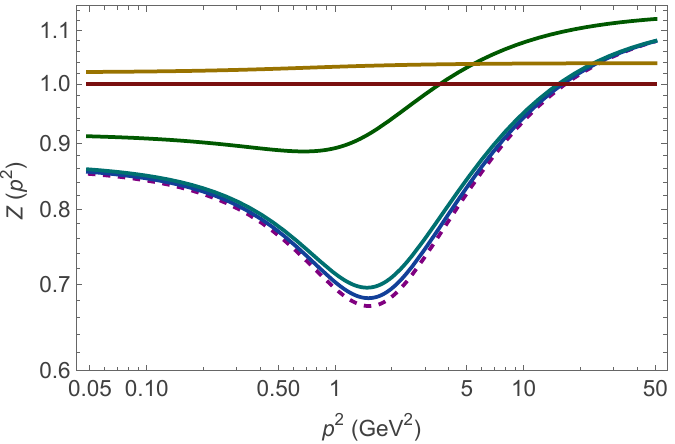}
	\vspace{2.2in}

	 \caption{(Color online) Predictions for the Model 2 quark wave function renormalizations from fitting the lattice data of Ref.~\cite{Bowman:2005vx} (left panels) and Ref.~\cite{Oliveira:2018lln} (right panels), in the chiral limit and for all 5 flavors in whole Minkowski space (top panels), the spacelike (middle panels), and timelike regions (bottom panels). Curves with stronger curvature correspond to
lighter quark mass, with the color code specified in Table~\ref{tab:masses}.}
	\label{fig:Zother2}
\end{figure*}

 The results for Model 2 are given in Table \ref{tab:masses}, and the corresponding quark mass functions in Figs.~\ref{fig:massother} and~\ref{fig:massother2}, and the corresponding quark $Z$ functions in Fig.~\ref{fig:Zother2}. Note that for very large masses (charm and bottom quarks) the quark mass and quark $Z$ functions are nearly constant. 
 
\begin{figure}
 	\includegraphics[height=1.9in]{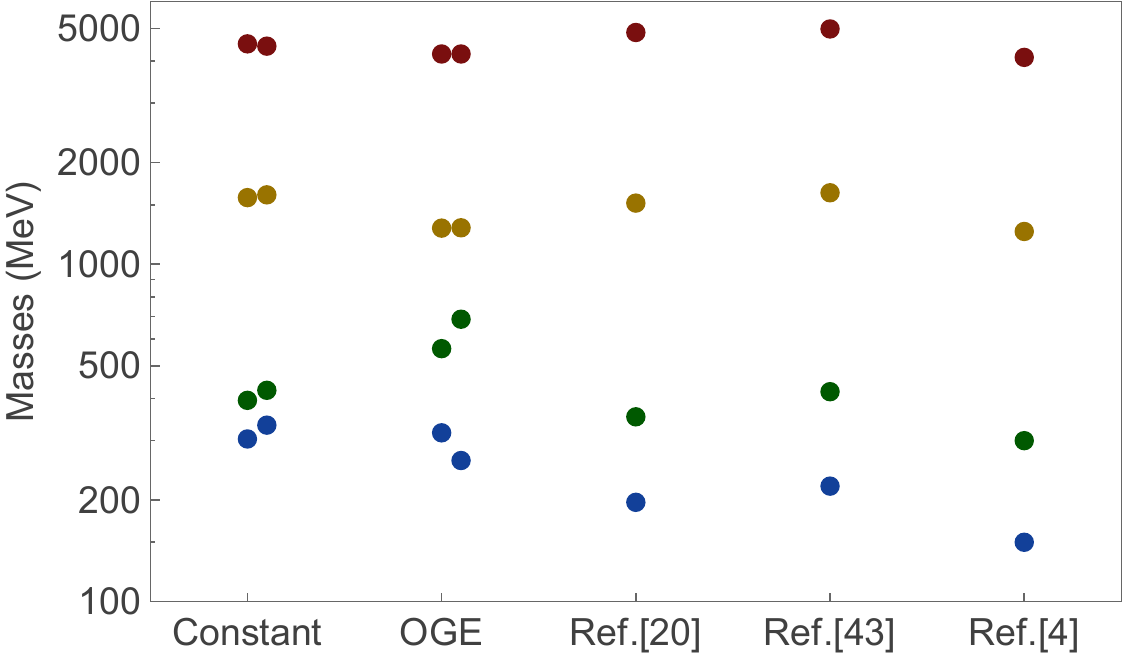}
 	 \caption{(Color online)  Comparison of constituent quark masses obtained in the paper with masses used in other calculations.  The five cases shown are discussed in the text. The dots are $m_{q}=\sfrac{1}{2}(m_{\rm u}+m_{\rm d})$, $m_{\rm s}$, $m_{\rm c}$, and $m_{\rm b}$, with $m_{q}$ colored darkblue and the others colored as as given in Table~\ref{tab:masses}.}

 	\label{fig:Massescom}
 \end{figure}

Figure \ref{fig:Massescom} compares the dressed (constituent) quark masses obtained in this paper with other calculations.  The two columns labeled "Constant" are masses obtained from the constant potential shown in Table \ref{tab:masses} (with the points slightly to the left obtained in the model fit to the lattice data of Ref.~\cite{Bowman:2005vx}, and to the right from Ref.~\cite{Oliveira:2018lln}), the two columns labeled "OGE" are obtained from the OGE potential (with the two columns as described above), and the other three columns are from Refs.~\cite{Stadler:2026acl}, \cite{Godfrey:1985xj}, and Sec.~9.4 in Ref.~\cite{Gross:2022hyw}.
Our masses are within the range of these other models, with the OGE results giving slightly larger values of $m_q$ and $m_{\rm s}$.

The final step in the renormalization procedure will now be discussed.

\subsection{Renormalization: eliminating the subtraction terms}\label{secIIIE}

The subtraction terms can be reabsorbed into a redefinition of the ``constant'' interaction discussed in Sec.~\ref{sec:3}. Collecting the subtraction terms together 
\bea
C_{\rm s}(p^2) &=&
\Big[{\overline A} (p^2)-A^{\rm S} (p^2)\Big] 
\nonumber\\ &&+\Big[A^{\rm S} (p^2)-A^{\rm R} (p^2)\Big]
\nonumber\\
C_{\rm v}(p^2) &=& \Big[{\overline B} (p^2)-B^{\rm S} (p^2)\Big] 
\nonumber\\&&+\Big[B^{\rm S} (p^2)-B^{\rm R} (p^2)\Big]\, , \qquad\quad
  \label{eq:Sub}
\eea
then the renormalized result can be expressed as the difference between the original singular result and the subtration terms 
\bea
A ^{\rm R}(p^2)&=&\overline{A} (p^2)-C_{\rm s}(p^2)\nonumber\\
B ^{\rm R}(p^2)&=&\overline{B} (p^2)-C_{\rm v}(p^2) \, .
\eea
But these subtraction terms, which depend on $p^2$, have the form of the ``constant" terms $C_{\rm s}$ and $C_{\rm v}$ that were discussed in Sec.~\ref{sec:3} above. 
Hence if these terms are added to any constant interaction used, they will cancel the singular parts of the OGE contribution, leaving both finite.  This is symbolized by the equation
\bea 
{\cal V}_{\rm g}(\hat{q}^2)\big|_{\rm sing}+{\cal V}_{\rm c}\big|_{\rm sing}={\cal V}_{\rm g}^{\rm R}(\hat q^2)+ {\cal V}_{\rm c}\, ,\qquad
\eea
where $ {\cal V}_{\rm c}$ will be zero in models with an OGE mechanism only.

\subsection{How mass functions modify $q\bar{q}$ calculations} \label{sec:masslimit}

The masses of $q\bar{q}$ mesons are calculated from the Gross equation (see Fig.~\ref{fig:BSVertexconst} from Ref.~\cite{Leitao:2017mlx}).  To find out how the the quark mass functions modify calculations using the one-channel Gross equation (where on-shell contributions come only from the quark) examine the structure of the diagram where a quark of mass $m_1$ is on shell and a quark with mass $m_2$, moving in the $-k_2$ direction, is off shell:
\bea
&&\Gamma(\hat{p}_1,P) =  -Z_{m_1} Z_{m_2}\int \frac{\mathrm d^3k_1}{ (2\pi)^3} \frac{m_1}{E_{k_1}} {\cal V}(q^2) \frac{m_2-\slashed{k}_2}{m_2^2-{k}_2^2-\rm i\, \epsilon } \nonumber\\&&\qquad\qquad\times\Gamma(\hat{k}_1,P) \Lambda(\hat k_1)\, \nonumber\\
&&\to - Z_{m_1}\int \frac{\mathrm d^3k_1}{ (2\pi)^3}  \frac{m_1}{E_{k_1}} {\cal V} (q^2) Z_{2}(k_2^2)\frac{M_2({k}_2^2)-\slashed{k}_2}{M_2^2({k}_2^2)-k_2^2-\rm i\, \epsilon } \nonumber\\&&\qquad\qquad\times\Gamma(\hat{k}_1,P)  \Lambda(\hat k_1) \,.\nonumber\\
\eea 
Here, in the second line, the mass function  $M_2({k}_2^2)$ and wave-function renormalization $Z_2({k}_2^2)$  of quark 2 replace the fixed mass $m_2$, but the mass of the on-shell quark with four-momentum $\hat k_1=(E_{k_1},{\bf k}_1)$ is fixed because it is evaluated at the mass pole.     The dressed antiquark mass depends on its four-momentum, and in the rest frame of the bound state this is bounded from {\it above\/} by
\bea
 k_2^2&=&(\hat k_1 - P)^2 = m_1^2 +M_b^2 -2 E_{k_1} M_b\nonumber\\
&\leq& (m_1-M_b)^2\, . \qquad \label{eq:p2max}
\eea
For light mesons ($\pi, K, \eta, \rho, \omega$) these limits  are all less than  
0.3 GeV$^2$, and provide extra convergence.

The situation is similar for the two- and four-channel Gross equation, where the off-shell contributions from quark 1 also contribute. 
In this case the kernel is evaluated at the {\it negative energy pole} of the antiquark $m_2$, so that $\hat k_2 = (-E_{k_2}, {\bf k}_2)$, and ${k}^2_1$ is bounded from {\it above} by
\bea
 k_1^2&=&(\hat k_2 + P)^2 = m_2^2 +M_b^2 -2 E_{k_2} M_b\nonumber\\
&\leq& (-m_2+M_b)^2\, . \qquad \label{eq:p1max}
\eea
The three versions of the Gross equation therefore give similar corrections.

\section{Discussion and conclusions} \label{sec:IV}

\subsection{Assessment of this work}

{Within the Covariant Spectator Theory, the dressed quark mass functions and quark $Z$-functions can be
calculated in the timelike region of $p^2$. In this work, we constrain models of the quark mass
function by imposing the following conditions: (i) consistency with lattice-QCD results for the
dressed OGE propagator; (ii) consistency with lattice-QCD results for the
dressed quark propagators obtained by two independent groups; (iii) the requirement that the
mass shell equation,
$
M(m^2,m_0)=m,
$
has one and only one solution; and (iv) the requirement that the quark mass function,
$
M(p^2,m_0)
$,
is a monotonically increasing function of the bare quark mass $m_0$. Of the three models considered, only Model~2 satisfies all of these constraints. 
By fitting three parameters to each lattice-data set associated with a given lattice bare quark mass, 
we obtain predictions for the dressed masses of the five physical quarks studied here: up, down, strange, charm, 
and bottom. The top quark is not considered. The results are summarized in Table~\ref{tab:masses}.

We consider lattice results from two independent lattice groups. 
The fits to these independent mass-function data sets yield similar predictions, supporting the stability of the model. Overall, the
predicted masses are reasonable, although the strange-quark mass, comes out larger than
expected. The Model~2 fits to the lattice mass functions are very good for
the two lightest quarks in each data set. For the two heaviest lattice quarks, the agreement
is less precise, especially in one of the two data sets, suggesting a possible discrepancy between the two data sets and/or directions for
future refinements of the model. In hindsight, the strange quark sector makes the transition from the light relativistic quark domain, where chiral symmetry breaking is essential, to the heavier non-relativistic quark domain. On the other hand, by construction, CST is specially tailored for the relativistic behavior of the light-light meson sector and chiral symmetry constraints, as well as for heavy-light systems where the heavy particle energy pole is dominating.

The main conclusion of this work is that it provides the proof of principle that it is possible to
construct a model for the quark mass function in Minkowski space that satisfies all required physical and lattice-QCD constraints while
also producing realistic predictions for the physical quark masses. Moreover, the
present framework establishes a foundation for developing improved
models that fit the lattice data more accurately, translating these data into knowledge of the quark-quark interaction kernel, and yield more realistic predictions, especially for the dressed strange-quark mass.

\begin{figure}
\includegraphics[width=0.45\textwidth]{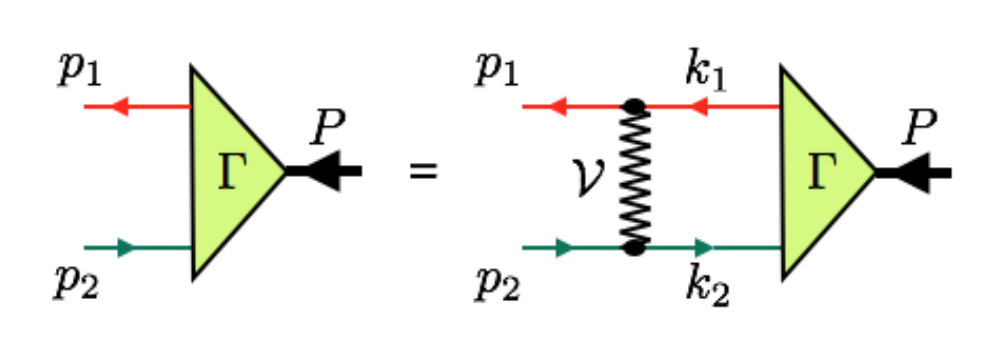}
 \caption{(Color online) {\it (Taken from Ref.~\cite{Leitao:2017mlx}.)\/} The BS equation for a quark-antiquark bound state. The bound state with four-momentum $P$ and external quark with four momentum $p_1$ (red line)  are moving to the left (forward in time), while the antiquark (green line) is represented as a quark moving backward in time with momentum $p_2$ (or an antiquark moving forward in time with physical momentum $-p_2$).   If the the external and internal legs are placed on shell in the correct manner, one obtains either one-channel, two-channel, or four-channel Gross equations used in CST to calculate the bound state. }\label{fig:BSVertexconst}
\end{figure}

\subsection{ Implications for future work}

 \begin{figure}[b]
    \centering
\includegraphics[height=2in]{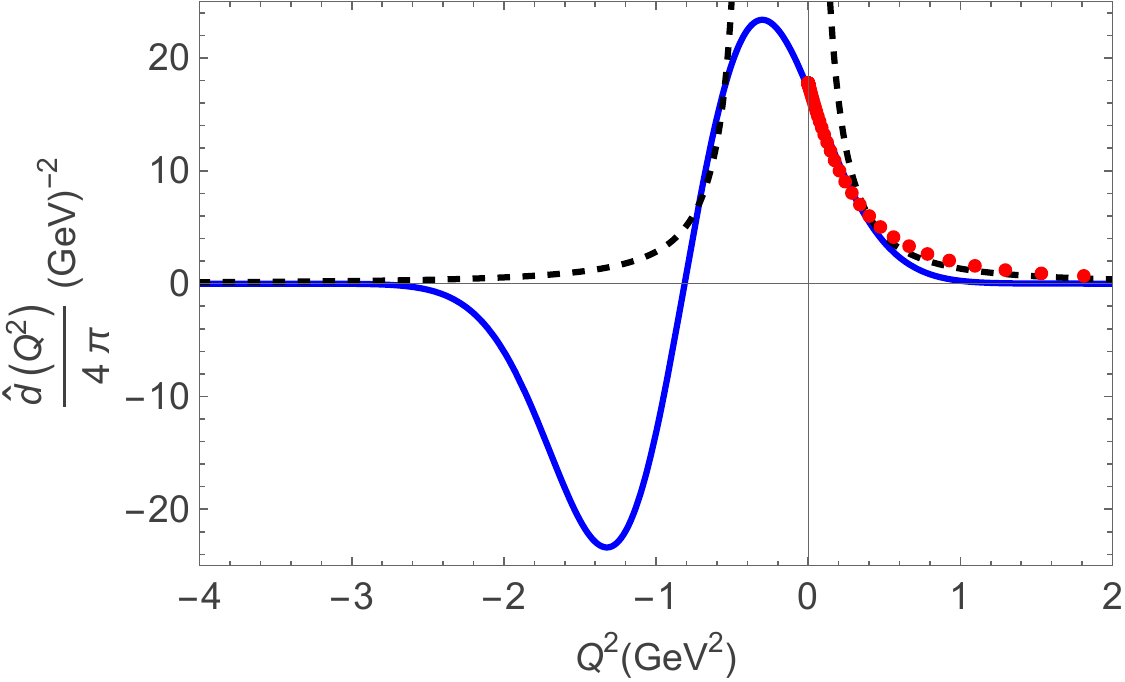}
 \caption{(Color online) The gluon propagator from Ref.~\cite{Leitao:2017mlx} (black dashed line) compared to Model 2 (blue line) and the lattice data.  Compare with Fig.~\ref{fig:D00}.} 
\label{fig:D16}
\end{figure}

The successful calculation of the meson spectrum given in Ref.~\cite{Leitao:2017mlx} uses a $q\bar{q}$ interaction kernel consisting of a confining interaction, an OGE interaction (in Feynman-'t Hooft gauge), and a scalar constant interaction  of the type introduced in Eq.~(\ref{eq:Ckernel}). 

This paper does not include a confining interaction. As discussed in Ref.~I, only a vector component of the confining interaction can contribute to the self-energy, and since  
 good fits to the meson spectrum are possible without such a vector component, assuming it is zero is not a serious limitation.

The OGE interaction used in Ref.~\cite{Leitao:2017mlx} is
\bea
V_{\rm OGE}=-4\pi \alpha_{\rm s}\Big(\frac1{\hat{q}^2}+\frac1{\Lambda^2-\hat{q}^2}\Big)\, ,
\eea
 with the choice $\Lambda=2m$ and different fits giving $\alpha_{\rm s}$ varying from 0.361 to 0.417.    Choosing  $\alpha_{\rm s}=0.4$ and $m=m_\chi = 0.3$ GeV, this form is compared with Model 2 in Fig.~\ref{fig:D16}. 
 Note that it agrees well with the lattice data for $Q^2\gtrsim 0.6$ GeV$^2$, so redoing the fits using Model 2 may be possible.

To sum up: the major development in this paper is the introduction  of a new way to regularize the infinities that emerge when the lattice data for the OGE is used to calculate the invariant  quark mass functions directly in Minkowski space. The results can be used in calculations of the meson spectrum.  Subtractions can be found that make $A$ and $B$ {\it finite and continuous\/} at the troublesome 
transition region in the vicinity of $p^2=0$.

 \vspace{1in}
\begin{acknowledgments}
	This work was supported by FCT under grant numbers CERN/FIS-PAR/0023/2021 and  UID/04349/2025 (https://doi.org/10.54499/UID/04349/2025). 
\end{acknowledgments}

\begin{appendix}

\section{Analytic results} \label{App:A}

Denoting $m^2 + p^2\equiv R$ and $2 \beta -\gamma \equiv T $, and recalling that  the error function is
 \bea
 {\rm erf}(x) = \frac{2}{\sqrt{\pi}}\int_0^x \exp [-t^2]\,\mathrm dt\, ,
 \eea
 the integrals Eq.~(\ref{eq:smooth}) have the form (\ref{eq:smooth2}) with
\begin{widetext}
\begin{subequations}
\bea
a ^{\rm S}(p^2) &&= \frac{M_1^2}{2 \beta } \left\lbrace\sqrt{\pi } {\rm e}^{\frac{\gamma ^2}{4}} \text{erf}\left(\frac{R}{M_1^2}-\frac{\gamma }{2}\right) \left[2  T R+M_1^2 \left(2-\gamma  T\right)\right]+2 M_1^2  T\exp\left[-\frac{R \left(R-\gamma  M_1^2\right)}{M_1^4}\right]\right\rbrace
     \label{eq:ai} \\
 b ^{\rm S}(p^2)&&= \frac{M_1^2}{16 \beta  p^2} \left\lbrace\sqrt{\pi } {\rm e}^{\frac{\gamma ^2}{4}} \text{erf}\left(\frac{R}{M_1^2}-\frac{\gamma }{2}\right) \bigg[-4 \beta  \left(2 R^2+M_1^4\right)-2 \gamma ^2 M_1^2 \left(2 R+\beta  M_1^2\right)+4 \gamma  R \left(R+2 \beta  M_1^2\right)\right.\nonumber\\&&\left.-8 M_1^2 R+\gamma ^3 M_1^4+6 \gamma  M_1^4\bigg]-2 M_1^2 \exp\left[-\frac{R \left(R-\gamma  M_1^2\right)}{M_1^4}\right] \left[2  T R+M_1^2 \left(4-\gamma  T\right)\right]\right\rbrace \, .\label{eq:bi}
 \eea
 \end{subequations}
 \end{widetext}
 Note that $b^{\rm S}(p^2) $ has a pole at $p^2\to 0$, so that $B ^{\rm S}$ behaves like a dipole at $p^2=0$.

To regularize $A^{\rm S}$ it is sufficient to subtract $a $ at $p^2 = 0$, giving 
 \bea
 a ^{\rm R}(p^2)= a ^{\rm S}(p^2)-a ^{\rm S}(0)\, .
 \eea
 Then, 
 \bea
 \lim_{p^2\to\pm \infty}\frac{a ^{\rm R}(p^2)}{p^2} = \mp\frac{ {\rm e}^{\frac{\gamma^2}{4}}\sqrt{\pi}\,M_1^2 T}{\beta } \, . \label{eq:asym}
 \eea
Therefore, the requirement that 

\bea
A^{\rm R}(p^2)=\frac{ Z_m\alpha_0}{3 \pi p^2}a^{\rm R}(p^2)
\eea 
approaches zero as $p^2$ approaches $\pm\infty$ gives $ T=0$, or
\bea
\beta  =\frac{\gamma}{2}\, . \label{eq:gamma}
\eea
 
 This is a significant constraint on the models, explaining why the choices of $\gamma=0$ or $\beta =\infty$ [the absence of a polynomial term of the form $(1 - q^2/\beta M_1^2)$] will not work.

To regularize $B^{\rm S}$, expand $b (p^2)$ in a power series at $p^2=0$.  The first two terms are
\bea
b (p^2)\to \frac{b _{-1}}{p^2} + b_0 + {\cal O}(p^2)
\eea
with
\begin{widetext}
\bea
&&b_{-1}=\lim_{p^2\to 0} p^2 b (p^2)\nonumber\\
&&b_0= -\frac{1}{16 \beta }\left\lbrace4 \sqrt{\pi } {\rm e}^{\frac{\gamma ^2}{4}} M_1^2 \text{erf}\left(\frac{m^2}{M_1^2}-\frac{\gamma }{2}\right) 
\left[2 m^2  T+M_1^2 \left(2-\gamma  T\right)\right]-8 M_1^4 \left(\gamma -2 \beta \right) \exp\left(\frac{\gamma  m^2 M_1^2-m^4}{M_1^4}\right)\right\rbrace\qquad
\eea
\end{widetext}
so the regularized form of $b $ is
\bea
b ^{\rm R}(p^2)=b (p^2)- \frac{b_{-1}}{p^2} - b_0\, . 
\eea
It is left to the reader to show that condition (\ref{eq:gamma}) also ensures that 
\bea
B^{\rm R}(p^2)=\frac{Z_m\alpha_0}{3 \pi p^2}b^{\rm R}(p^2)
\eea
vanishes as $p^2$ goes to $\pm\infty$.

\section{Comparison with Ref.~I} \label{App:B}

There are many differences between this work and Ref.~I. The two major differences are summarized here:

\begin{itemize}
  
  \item The model gluon mass functions chosen in this paper converge at large values of $k$ (or $y$). The only singularities that remain are at $p^2=0$.  This simplifies the calculations and removes the need for the form factor $g(y)$ (where $y^2=\frac{(k\cdot p)^2}{k^2p^2}$ in Ref.~I) that does not depend solely on $q^2$.
  
  \item To regularize the integrals, Ref.~I made the replacement  $q^2\to -|q^2|$.  This does not give the desired cancellation of the $E_k p_0$ factors that leads directly to the observation that all physical quantities depend only on $p^2$ and not on $p_0$.  In this paper the models can be integrated analytically, and shown to depend only on $p_0^2\to p^2$.   
\end{itemize}

\end{appendix}

\bibliography{PapersDB-v2-4}

\end{document}